\documentclass[aip,rsi,reprint,graphicx,amsmath,amssymb,twocolumn]{revtex4-1}
\usepackage{graphicx}
\usepackage{dcolumn}
\usepackage{bm}
\usepackage{color}
\usepackage{gensymb}
\begin{document}
\title{An apparatus for studying electrical breakdown in liquid helium
  at 0.4 K and testing electrode materials for the SNS nEDM experiment} 

\author{T.~M.~Ito}
\email[Corresponding author. Email: ]{ito@lanl.gov}
\affiliation{Los Alamos National Laboratory, Los Alamos, New Mexico 87545, USA}

\author{J.~C.~Ramsey}
\affiliation{Los Alamos National Laboratory, Los Alamos, New Mexico 87545, USA}

\author{W. Yao}
\affiliation{Oak Ridge National Laboratory, Oak Ridge, Tennessee
  37831, USA}

\author{D.~H.~Beck}
\affiliation{Loomis Laboratory of Physics, University of Illinois,
  Urbana, Illinois 61801, USA}

\author{V.~Cianciolo}
\affiliation{Oak Ridge National Laboratory, Oak Ridge, Tennessee
  37831, USA}

\author{S.~M.~Clayton}
\affiliation{Los Alamos National Laboratory, Los Alamos, New Mexico 87545, USA}

\author{C. Crawford}
\affiliation{Department of Physics and Astronomy, University of
  Kentucky, Lexington, Kentucky 40506, USA}

\author{S.~A.~Currie}
\affiliation{Los Alamos National Laboratory, Los Alamos, New Mexico 87545, USA}

\author{B.~W.~Filippone}
\affiliation{W.~K.~Kellogg Radiation Laboratory, California Institute of
  Technology, Pasadena, California, 91125, USA}

\author{W.~C.~Griffith}
\affiliation{Los Alamos National Laboratory, Los Alamos, New Mexico 87545, USA}

\author{M.~Makela}
\affiliation{Los Alamos National Laboratory, Los Alamos, New Mexico 87545, USA}

\author{R. Schmid}
\affiliation{W.~K.~Kellogg Radiation Laboratory, California Institute of
  Technology, Pasadena, California, 91125, USA}

\author{G.~M.~Seidel}
\affiliation{Department of Physics, Brown University, Providence,
  Rhode Island, 02912, USA }

\author{Z.~Tang}
\affiliation{Los Alamos National Laboratory, Los Alamos, New Mexico 87545, USA}

\author{D. Wagner}
\affiliation{Department of Physics and Astronomy, University of
  Kentucky, Lexington, Kentucky 40506, USA}

\author{W.~Wei}
\affiliation{Los Alamos National Laboratory, Los Alamos, New Mexico 87545, USA}

\author{S.~E.~Williamson}
\affiliation{Loomis Laboratory of Physics, University of Illinois,
  Urbana, Illinois 61801, USA}

\date{\today}

\begin{abstract}
We have constructed an apparatus to study DC electrical breakdown in
liquid helium at temperatures as low as 0.4~K and at pressures between
the saturated vapor pressure and $\sim$600~torr. The apparatus can
house a set of electrodes that are 12~cm in diameter with a gap of
$1-2$~cm between them, and a potential up to $\pm 50$~kV can be
applied to each electrode. Initial results demonstrated that it is
possible to apply fields exceeding 100~kV/cm in a 1~cm gap between two
electropolished stainless steel electrodes 12~cm in diameter for a
wide range of pressures at 0.4~K. We also measured the current between
two electrodes. Our initial results, $I<1$~pA at 45~kV, correspond to
a lower bound on the effective volume resistivity of LHe of $\rho_V >
5\times10^{18}$~$\Omega\cdot$cm. This lower bound is 5 times larger
than the bound previously measured. We report the design,
construction, and operational experience of the apparatus, as well as
initial results.
\end{abstract}


\maketitle 

\section{Introduction}
A nonzero permanent electric dipole moment (EDM) of a nondegenerate
state of a system with spin $J\ne0$ violates invariance under time
reversal as well as invariance under parity operation. The
violation of time reversal invariance implies a violation of
invariance under $CP$ operation (combined operations of parity and
charge conjugation) through the $CPT$ theorem.

Searches for new sources of $CP$ violation through searches of EDM are
strongly motivated for the following reasons: (1)~The amount of $CP$
violation contained in the Standard Model (SM) of particle physics
through the complex phase of the Cabibbo-Kobayashi-Maskawa (CKM)
matrix is known to be insufficient for generating the observed
matter-antimatter asymmetry in the Universe (therefore it is expected
that there are other sources of $CP$ violation). (2)~The other source
of $CP$ violation contained in the SM, the so-called $\theta$ term in
the quantum chromodynamics (QCD) Lagrangian, is constrained by
nonobservation of the neutron EDM (nEDM) to be unnaturally
small. (3)~The SM value of EDMs based on the CKM phase are highly
suppressed because in the SM the $CP$ violation processes are quark
flavor-changing at the tree level. (4)~Many extensions of the SM
contain new sources of $CP$ violation, predicting much larger values
of EDMs.

The search for the neutron EDM (nEDM) was pioneered by Smith, Purcell,
and Ramsey.\cite{SMI57} In the subsequent 60 odd years, many more
experiments have been performed with increasingly refined experimental
methods and with correspondingly improved sensitivities and upper
limits. (For a review of experimental searches for the nEDM, see {\it
  e.g.} Ref.~\onlinecite{LAM09}.) Many of these experiments have
looked for the nEDM by subjecting spin polarized neutrons to static
magnetic and electric fields and looking for the possible change in
the spin precession frequency corresponding to the change in the
relative orientation of the electric field with respect to the
magnetic field. Whereas early experiments were performed using a beam
of cold neutrons, all recent experiments have used stored ultracold
neutrons (UCNs)\cite{IGN90,GOL91} to suppress the effect of the
motional magnetic field\cite{LAM09} (see, however,
Ref.~\onlinecite{PIE13}).

The statistical sensitivity of such measurements for a batch of
stored UCNs depends on three quantities, namely, $E$ the strength of the
electric field, $T$ the free precession time, and $N$ the number of
neutrons in the batch, and is expressed as follows:
\begin{equation}
\label{eq:sensitivity}
\delta d_n \propto
  \frac{1}{ET\sqrt{N}}.
\end{equation}
Typically many such measurements are repeated over the duration of an
experiment. The current limit, given by an experiment performed at
Institut Laue Langevin by a group led by the University of Sussex, is
$d_n < 2.9\times 10^{-26}$~$e\cdot$cm (90\% C.L.).\cite{BAK06} (A
recent updated analysis gives $d_n < 3.0\times 10^{-26}$~$e\cdot$cm
(90\% C.L.).\cite{PEN15}) In this experiment, typical values for $E$,
$T$, and $N$ were $E=10$~kV/cm, $T=130$~s, and $N=14,000$.

A new nEDM experiment, to be mounted at the Fundamental Neutron
Physics Beamline (FnPB)\cite{FOM15} at the Spallation Neutron Source
(SNS) at Oak Ridge National Laboratory, is currently being developed
with a sensitivity goal of $\delta d_n \sim 3 \times
10^{-28}$~$e\cdot$cm.\cite{nEDM,ITO07} This experiment is based on the
method proposed by Golub and Lamoreaux.\cite{GOL94} In this method,
the experiment is performed inside a bath of liquid helium (LHe) at
approximately 0.4~K. The unique features of this experiment include:
\begin{enumerate}
\item In-situ production of UCNs inside the measurement cells
from a cold neutron beam of 0.89~nm wavelength using the superthermal
process in superfluid liquid helium.\cite{GOL77}
\item Use of spin-polarized $^3$He atoms as a comagnetometer.
\item Use of the spin-dependent neutron capture reaction on a $^3$He
  atom ($n+^3{\rm He}\to p+t$) and resulting LHe scintillation (see
  {\it e.g.} Ref.~\onlinecite{ITO12}) as the analyzer of the neutron
  spin.
\end{enumerate}
With this method, we expect to increase all of $E$, $T$, and $N$ in
Eq.~(\ref{eq:sensitivity}) significantly over the previous
experiments. LHe is expected to be a better insulator than vacuum. In
addition, processes responsible for electrical breakdown initiated at
the electrode-insulator junction, thought to be one the factors
limiting the achievable electric field strength for previous room
temperature nEDM experiments, are expected to be suppressed in LHe.
Therefore a larger $E$ can be expected. Performing a measurement at
cryogenic temperatures suppresses some of the loss mechanisms for
stored UCNs, resulting in longer $T$.\cite{AGE85,KOR04} Producing UCNs
directly in the experiment from a 0.89~nm cold neutron beam is
expected to provide a larger $N$.

Of critical importance to the development of this experiment is to
establish the highest electric field that can be applied and be
sustained stably in the volume inside the measurement cells because
the sensitivity of the experiment directly depends on the strength of
the applied electric field.

As part of the research and development (R\&D) for the SNS nEDM
experiment, we have constructed an apparatus to study DC electrical
breakdown in LHe at temperatures as low as 0.4~K and at pressures
between the saturated vapor pressure (SVP) and $\sim$600~torr. In this
paper, we describe the design, construction, and operational
experience of this apparatus. We also report the initial findings. 

The rest of this paper is organized as follows. In
Sec.~\ref{sec:requirements}, the electric field requirements for the
SNS nEDM experiment will be reviewed. After a brief review of the
current understanding of phenomenon of electrical breakdown in LHe in
Sec.~\ref{sec:EBinLHe}, the purpose of the new high voltage (HV)
testing apparatus will be discussed and an overview of its design will
be given in Sec.~\ref{sec:DesignOverview}. The details of the
cryogenic design of the apparatus will be given in
Sec.~\ref{sec:CryoDesign}, whereas the design of the HV components
will be discussed in Sec.~\ref{sec:HVDesign}. The cryogenic
performance of the apparatus, as well as our operational experience
with cryogenic aspects of the apparatus, will be discussed in
Sec.~\ref{sec:CryoPerformance}. The HV performance, our operational
experience with HV aspects of the apparatus, and some initial HV
results will be presented in
Sec.~\ref{sec:HVPerformance}. Section~\ref{sec:Summary} will give a
summary of this paper.

\section{Electric field requirements for the SNS nEDM experiment}
\label{sec:requirements}
The design goal for the SNS nEDM experiment is to have an electric
field of 75~kV/cm stably applied in the volume inside the so-called
measurement cells, the volumes filled with 0.4~K LHe that stores
UCNs. The measurement cell walls will be made of
poly(methyl methacrylate) (PMMA), as they will serve as part of the
light collection system. They will be 10.16~cm$\times$ 12.70~cm
$\times$ 42~cm in outer dimension with a wall thickness of 1.2~cm. The
two measurement cells will be sandwiched between electrodes, roughly
$10$~cm~$\times$~40~cm$\times$~80~cm in size. The electrodes and the
measurement cells will be immersed in 0.4~K LHe.

There are a number of requirements on the electrode materials. They
are:
\begin{enumerate}
\item In order to match the thermal contraction of the measurement
  cells made of PMMA, which shrinks $\sim$1\% when cooled from room
  temperature to 0.4~K, the electrodes need to be made of a material
  that has similar thermal contraction characteristics to PMMA. The
  current design is to use PMMA as the electrode substrate material
  and make the electrode surface conducting using methods including:
  (1)~coating with appropriate materials and (2)~implanting
  conducting materials into the surface layer. 
\item The material cannot have too high an electrical
  conductivity. This requirement comes from the requirement on Johnson
  noise on the superconducting quantum interference device
  (SQUID)-based magnetometer to measure the precession frequency of
  spin polarized $^3$He atoms and also from the requirement on joule
  heating from eddy currents due to the radio frequency (RF) field for
  dressed spin measurement.\cite{GOL94} The allowed surface
  resistivity is 100~$\Omega/\square < \sigma < 10^8$~$\Omega/\square$ at the
  operating temperature of $\sim$0.4~K.
\item The material should be non-magnetic. The static magnetic field
  in the region inside the measurement cells, which is approximately
  1~$\mu$T, needs to be uniform to $5\times 10^{-4}$ and needs to have
  field gradients smaller than 5~pT/cm in the direction of the
  static field and 10~pT/cm in the direction perpendicular to the
  static field. Because of this stringent requirement, many of the
  so-called ``non-magnetic'' technical materials, such as stainless
  steel and inconel, are disallowed. Also, materials that become
  superconducting cannot be used because the field expelled due to the
  Meissner effect would disturb the field uniformity inside the
  measurement cells.
\item The material should not have large neutron absorption
  properties, as such materials would become radioactively activated
  due to the exposure to a high flux neutron beam and become source of
  background radiation. 
\end{enumerate}
The current candidate methods for making the PMMA electrode surface
conducting include graphite paint and copper ion implantation.

In addition, the leakage currents along the cell walls need to be
minimized. This requirement comes from the following considerations.
\begin{itemize}
\item These currents produce magnetic fields that are correlated with
  the direction of the electric field and therefore can produce
  effects that mimic the signal of nEDM (this applies to all nEDM
  experiments that use stored UCN).
\item In the current design, the necessary HV will be generated inside
  the LHe volume using a gain capacitor. In this method, the nEDM
  experiment will be run with the HV electrode disconnected from the
  HV power supply. Leakage currents lead to a reduction of the
  electric field over time.
\item These currents produce heat, generating phonons in
  superfluid LHe and modifying the spatial distribution of
  the $^3$He atoms via the $^3$He-phonon interactions.
\end{itemize}

\begin{figure*}
\centering
\includegraphics[width=6in]{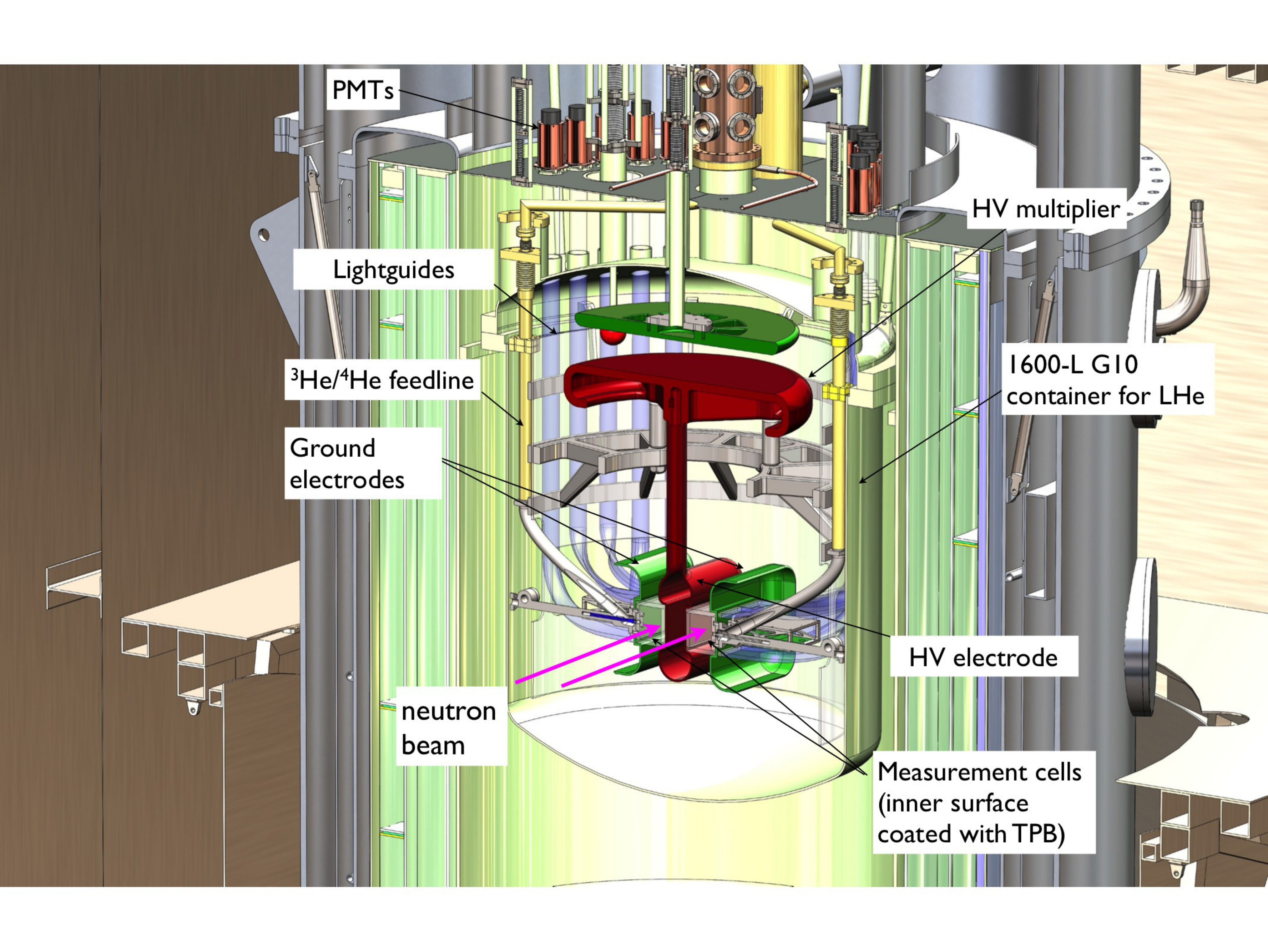}
\caption{A schematic of the central part of the SNS nEDM experiment,
  as it is currently designed\label{fig:CDSschematic}}
\end{figure*}

A schematic of the central part of the SNS nEDM experiment, as it is
currently designed, is shown in Fig.~\ref{fig:CDSschematic}.

\section{Electrical breakdown in liquid helium}
\label{sec:EBinLHe}
Electrical breakdown in liquid helium has been studied
extensively\cite{GER98} because of its importance in cooling
superconducting components, principally high field magnets. In
general, the experimental measurements of breakdown show little
consistency, varying with geometry, electrode material, bubble
formation and unexplained parameters.

An electron with low kinetic energy forms a bubble in liquid
helium. 
Because of the overlap with the electrons of the helium atoms,
a free electron has a lower energy by close to an eV compared to that
when it is in the conduction band. 
The radius of the bubble is about
19\AA. In a high electric field, an electron does not escape the
bubble. Its motion is constrained by viscous damping or in the
superfluid state by the creation of vortex rings or rotons, depending
on pressure. 
In contrast, because of electrostriction, positive He ions form
solid-like structures with surrounding atoms called ``snowballs.''
The kinetic energy of a snowball in an electric
field is also limited by damping and vortex ring creation. Neither
electron bubbles nor positive ion snowballs can be the source of
breakdown in liquid helium.

An electron created in the conduction band in a high electric field,
on the other hand, can in principle lead to breakdown by electron
impact ionization. However, the breakdown field of bulk liquid helium
is calculated to be very high. The mean free path of a conduction
electron, limited by elastic scattering from He atoms, is only about
100 nm and the ionization energy of He is 24.6
eV. Belevtsev\cite{BEL93} has made a detailed theoretical study of
electron multiplication induced by an electric field in liquid helium
starting from the kinetic Boltzmann equation. Taking into account the
various electron energy loss mechanisms, he finds that the first
Townsend coefficient reaches a value of 1~cm$^{-1}$ at a field of
1.5~MV/cm rising to 102~cm$^{-1}$ at 2.5~MV/cm. Hence, breakdown by
electron impact ionization in any reasonable sized volume of helium
can be expected to occur only at a field in the low MV/cm region, well
above the fields of interest for this experiment.  

Breakdown fields around or below~100 kV/cm have been commonly reported
in the literature for liquid helium. These observations are presumed
to be the result of phenomena occurring at the cathode. If the field
emission of electrons on a rough surface can occur because of
extremely high local fields at sharp asperities, local heating can
result in vapor bubble formation leading to electron multiplication
and vapor growth proceeding together. Breakdown is then the result of
a vapor column extending to the anode. Note that in other dielectric
liquids, formation and growth of vapor bubbles associated with
electrical breakdown have been observed
experimentally\cite{KAT89,KEL81,FOS90} (see also
Ref.~\onlinecite{RIZ14}).

It follows that the parameters that can affect the breakdown field
strength include: (1) electrode material, in particular the surface
properties and (2) LHe temperature and pressure (vapor growth can
depend on the pressure). Note that LHe is unique among liquids in that
its SVP changes drastically as a function of the temperature. For
example, the SVP is 760~torr at 4.2~K and is $\sim$$10^{-6}$~torr at
0.4~K.

In addition, because electrical breakdown is a stochastic process, the
size of the system affects the breakdown field strength and its
distribution, as shown in Ref.~\onlinecite{WEB56}, in which the
experimentally measured dependence of the breakdown field on the
electrode surface area was shown to agree with a prediction based on
the measured breakdown field distribution for electrodes of a
particular size and the theory of extreme values, for the case in
which transformer oil was used as insulator fluid.

In general, particles and other contamination in the liquid affect the
electrical breakdown properties.\cite{RIZ14} In the case of LHe,
however, it is expected that they play a less important role because
all possible contaminating species freeze at LHe
temperature. Additionally solid particles (made of {\it e.g.}  metal)
would most likely settle at the bottom of a container due to the low
density of LHe ($\rho \sim 0.145$~g/cm$^3$ for $T< 2$~K).

\section{Purpose of the apparatus and overview of the design}
\label{sec:DesignOverview}
The considerations given above indicate that the R\&D for the SNS nEDM
experiment requires a study of electrical breakdown in LHe in a
condition ({\it i.e.} temperature, pressure, and size) close to that
expected for the SNS nEDM experiment, using suitable electrode
candidate materials.

It is also very important to study the effect of the presence of a
dielectric insulator sandwiched between electrodes, as such will be
the geometry for the SNS nEDM experiment. Note that even in a room
temperature vacuum system, electric fields exceeding a few 100~kV/cm
are possible when there is no insulator directly sandwiched between
the two electrodes. In a study performed using a room temperature
vacuum apparatus\cite{GOL86} similar to those used in the previous
nEDM experiments (such as Ref.~\onlinecite{BAK06}), the electric field
was limited to $\sim$30~kV/cm due to the presence of the UCN confining
wall that was sandwiched between the two electrodes. (In
the actual nEDM experiment, the achievable field was further lowered
to $E\sim 10$~kV/cm due to other factors.)  Field emission at the
cathode-insulator junction is thought to be responsible for initiating
breakdown,\cite{KOF60} which is temperature independent. However, we
expect the processes responsible for leading field emission to
breakdown to be suppressed in LHe. 

In order to study the relevant aspects of electrical breakdown in LHe
with a goal of establishing the feasibility of the SNS nEDM experiment
as well as guiding the design of the apparatus, we constructed an
apparatus called the Medium Scale HV (MSHV) Test Apparatus, which is
described in this paper. 

There are some important design aspects that required close
attention to ensure that the apparatus yielded properly interpretable
results. The inconsistencies seen in previous results may well have
been partly due to lack of proper attention to these aspects.

One is proper electrostatic design. As discussed earlier, because
electrical breakdown is a stochastic process, the size of the system
affects the breakdown field strength and its distribution. It is
therefore important to have a good understanding of the area of the
electrodes that is exposed to high fields (and/or the size of the
stressed dielectric volume) for proper interpretation of the
results. For example, in a typical plane-to-plane geometry in which
electrodes are made of a flat disk whose edge is rounded with a
constant radius of curvature, it can be shown that the field is the
highest where the flat surface connects to the rounded edge. If the
distance between the two such electrodes is varied in order to study
the gap size dependence of the breakdown field, one can be led to
erroneous results because the ratio between the highest field and the
field at the center of the electrodes increases as the gap size is
increased. In order to study the breakdown field in a plane-to-plane
geometry, so-called uniform field electrodes need to be used.

Another is proper heat management. Due to the rather low heat of
vaporization of liquid helium (83~J/mol), a small amount of heat
transmitted to the electrodes through HV feed lines can boil LHe on
the electrode surfaces, creating bubbles and significantly reducing
the breakdown field. It is, therefore, of critical importance to
properly thermally anchor HV feed lines. Similarly, it is important to
make sure that all components in the system are properly cooled
before any measurements are taken. Performing breakdown field
measurements before a sufficient amount of heat is removed from all
the relevant components can lead to erroneous results due to boiling
of LHe caused by residual heat in the system.

We used a $^3$He refrigerator to cool the Central Volume (CV), the
volume that contains LHe and houses electrodes to be tested,
to the operating temperature of 0.4~K.  The size of the CV was
determined by striking a balance between the following two competing
factors:
\begin{itemize}
\item Short turnaround time (time from the beginning of one cooldown
  to the beginning of the next) of the system ($\sim$2~weeks) to allow
  multiple electrode material candidates to be tested.
\item Size large enough to give information relevant for the SNS nEDM experiment.
\end{itemize} 
The chosen size of the CV holds 6~liters of LHe, housing a set of
electrodes 12~cm in diameter. The gap size is adjustable between 1 and
2~cm. Each dimension is within a factor of 5 or so of the SNS nEDM
experiment's measurement cell electrodes.

Since we expect the pressure to be an important parameter affecting
the breakdown field strength in LHe, the MSHV system is designed so
that the pressure of the LHe volume in which the electrodes are placed
can be varied between the SVP and $\sim$600~torr. 

\begin{figure*}
\centering
\includegraphics[width=6in]{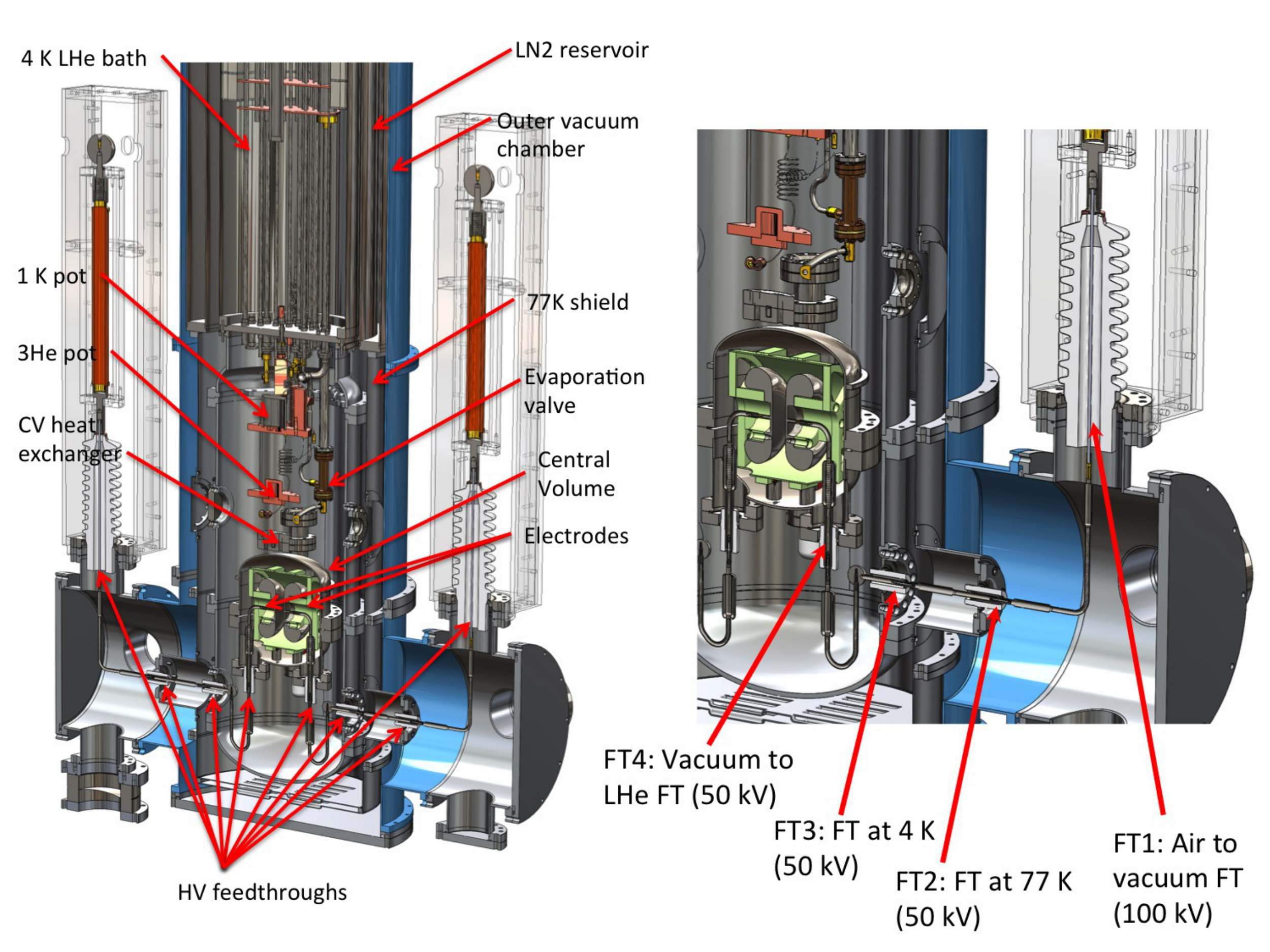}
\caption{A schematic of the MSHV system\label{fig:schematic}}
\end{figure*}

A schematic of the MSHV system is shown in Fig.~\ref{fig:schematic}.

\section{Cryogenic design}
\label{sec:CryoDesign}
As shown in Fig.~\ref{fig:schematic}, the MSHV cryostat consists of
the following major components:
\begin{itemize}
\item Outer vacuum chamber (OVC), which separates the insulating vacuum
  from the atmosphere.
\item LN$_2$ reservoir and the 77~K heat shield attached to it,
\item 4~K LHe bath.
\item Inner vacuum chamber (IVC), which also serves as a 4~K heat shield.
\item $^3$He refrigerator insert, which contains the $^3$He pot and
  the 1K pot.
\item the CV.
\item HV feedthroughs and interconnects.
\end{itemize}
The dimension of the OVC is 50.8~cm in diameter and 185~cm in
  height. The IVC is made of 1100 series Al alloy for its excellent
thermal conductivity. The IVC allows introduction of exchange gas to
help cool the $^3$He refrigerator insert and the CV in the process of
cooldown. 

\begin{figure}
\centering
\includegraphics[width=3.5in]{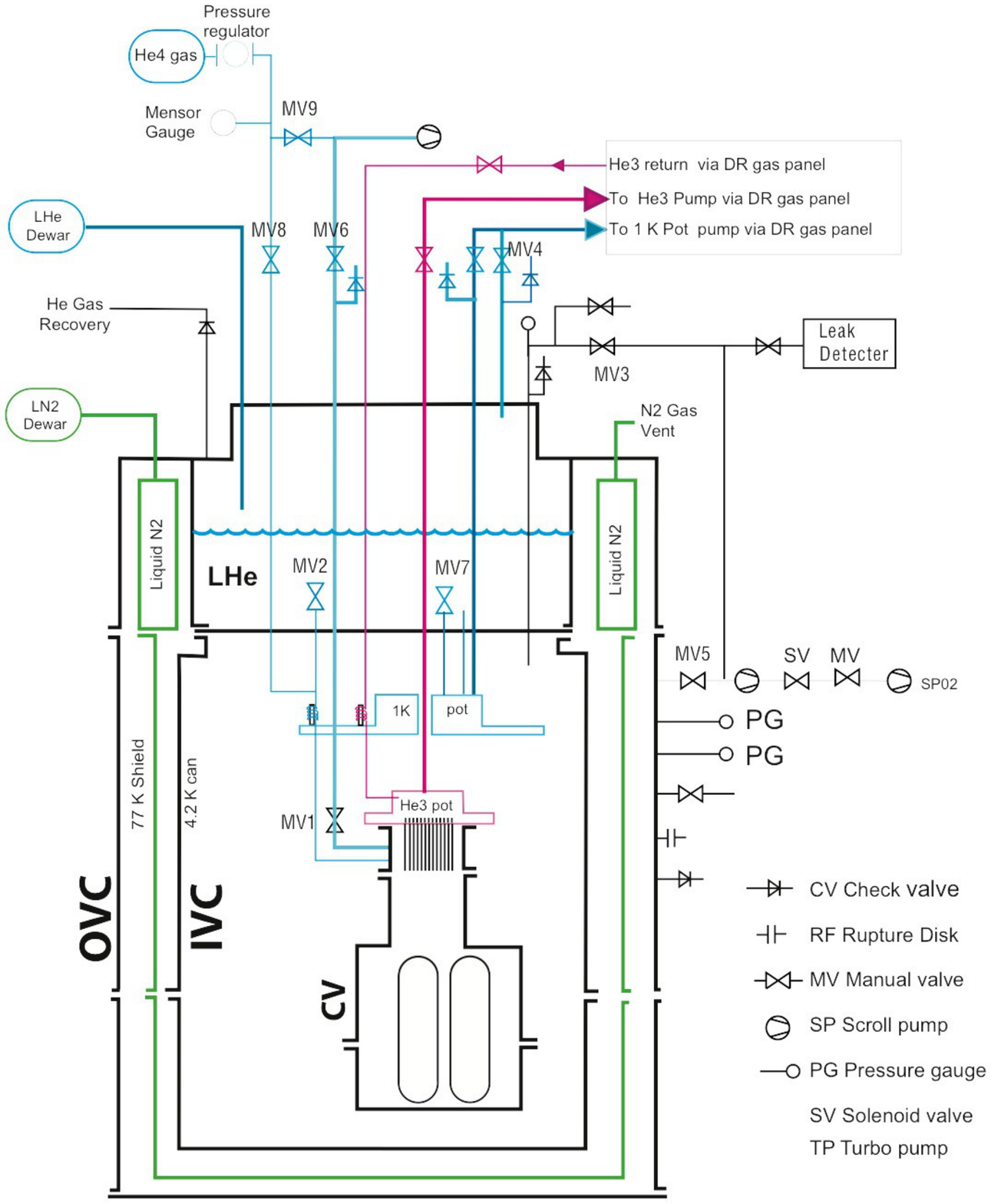}
\caption{MSHV plumbing diagram\label{fig:plumbing}}
\end{figure}

A diagram of the plumbing of the MSHV system is shown in
Fig.~\ref{fig:plumbing}.
The CV is cooled by the $^3$He pot through a heat exchanger mounted at
the bottom of the $^3$He pot. The heat exchanger is made by cutting
slots into a block of oxygen-free high thermal conductivity (OFHC)
copper with electrical discharge machining.

There are two lines connected to the CV. One is the ``fill'' line,
which connects the 4~K LHe bath to the CV through manual valve MV2,
located at the bottom of the 4~K LHe bath. The primary purpose of the
fill line is to introduce LHe into the CV. The fill line is thermally
anchored at the 1~K pot. The other is the ``pump out'' line, which
connects the CV to a vacuum pump through a large-aperture (6.35~mm)
superfluid-tight valve MV1.\cite{WIL15} The primary purpose of the
pump out line is to evacuate the CV prior to a cooldown of the system
and remove LHe from the CV during a warmup of the system.

Opening MV2 allows LHe to flow from the 4~K LHe bath to the
CV. Leaving it open keeps the pressure of the LHe inside the CV at the
pressure of the LHe inside the 4~K LHe bath, which is typically
600~torr, primarily determined by the atmopheric pressure in Los
Alamos, New Mexico, USA (elevation $\sim$2000~m). Note that there is a
temperature gradient established in the LHe column inside the CV fill
line, and the associated flow of heat to the CV, in this mode of
operation. Closing MV2 and pumping on the fill line to remove
LHe from the fill line allows the pressure in the CV to be
reduced. The pressure in the CV will be the SVP of LHe at the
temperature at the location of the liquid surface, which is somewhere
in the fill line. The diameter and length of the CV fill line were 
determined as a compromise between the following two competing
factors:
\begin{itemize}
\item Heat flow rate. During operation, when the CV fill line is
  filled with LHe, there is a continuous flow of heat through the LHe
  from the 1~K pot to the CV, which is typically operated at
  0.4~K. This must be sufficiently small so that it does not affect
  the temperature of the CV.
\item He flow rate. The CV needs to be filled within a reasonable
  amount of time.
\end{itemize}
We evaluated the heat flow rate from 1.5~K to 0.4~K through a
superfluid LHe column as a function of the diameter for selected
lengths according to Ref.~\onlinecite{BER68} using various
thermodynamic quantities of LHe found in
Ref.~\onlinecite{DON98}. Furthermore, we estimated the pressure drop
necessary to have a sufficient flow to fill the CV in 24 hours using
formulas from Ref.~\onlinecite{SNY92}. As a result of these
evaluations, we chose the following for the length and the diameter of
the fill line:
\begin{itemize}
\item LHe to 1~K pot: 47 cm long, 0.7~mm inner diameter
\item 1~K pot to CV: 76 cm long, 0.7~mm inner diameter
\end{itemize}

The pump out line is $\sim$6~mm in inner diameter. The
large-apertuture superfluid-tight valve is made of
polyamide-imide.\cite{WIL15} 
The valve was mechanically
actuated by using a rod, the other end of which is at room
temperature. In order to properly thermally anchor the actuation rod
to the 4~K shield, we used a solderless flexible thermal
link.\cite{SDL}

The CV, which houses the electrodes and holds about 6~liters of LHe,
is made of stainless steel. It is an ASME\cite{ASME} stamped vessel
with a maximum allowable working pressure of 8.6~bar (125~psi). The
vessel is also equipped with a cryogenic burst disk\cite{OSECO} with a
bursting pressure of 8.6~bar. This pressure rating of the CV was
determined based on the following safety considerations. If the
electricity to run the pumps for the $^3$He is lost during the
operation of the system, then the temperature of the CV will slowly
rise and will remain at $\sim$4~K as long as there is LHe in the 4~K
LHe bath, as the CV is inside the 4~K heat shield. If the system is
left unattended with MV2 closed, then the LHe inside the CV is
confined and the pressure inside the CV will rise as the temperature
of the CV rises. The pressure necessary to keep LHe in a fixed volume
as the temperature rises is calculated to be $\sim$7~bar (see
Appendix).

The CV can be opened at the 12-inch CF flange, allowing the electrodes
to be replaced. The 12-inch CF flange with standard copper gasket
provides a reliable superfluid LHe tight seal. 

In addition to several temperature sensors, both silicon diodes and
ruthenium oxide sensors, two LHe level sensors are installed inside
the CV. One is a capacitance level sensor, which is made of two
concentric cylinders mounted around the heat exchanger. This level
sensor allows monitoring of the LHe level in the CV when the CV is
close to being full. The dimensions of the cylinders are $\sim$7~cm in
height and $\sim$4.5~cm in diameter. The gap between the inner and outer
cylinders is $2.5\times 10^{-2}$~cm, giving a capacitance of
$\sim$300~pF. It is read out using a handheld LCR meter, Agilent Model
U1733C, which provides more than sufficient resolution and accuracy to
measure the change in capacitance caused by the dielectric constant of
LHe ($\sim$1.05). The other LHe level sensor is of the superconducting
wire type (American Magnetics Inc. Model 135-2K). This level sensor
provides the LHe level information during the initial fill and the
emptying of the CV. Because the heat imparted to LHe when operating
this sensor far exceeds the cooling power of the $^3$He refrigerator,
we keep this sensor turned off during normal operation.

\begin{figure}
\centering
\centering
\includegraphics[width=3.5in]{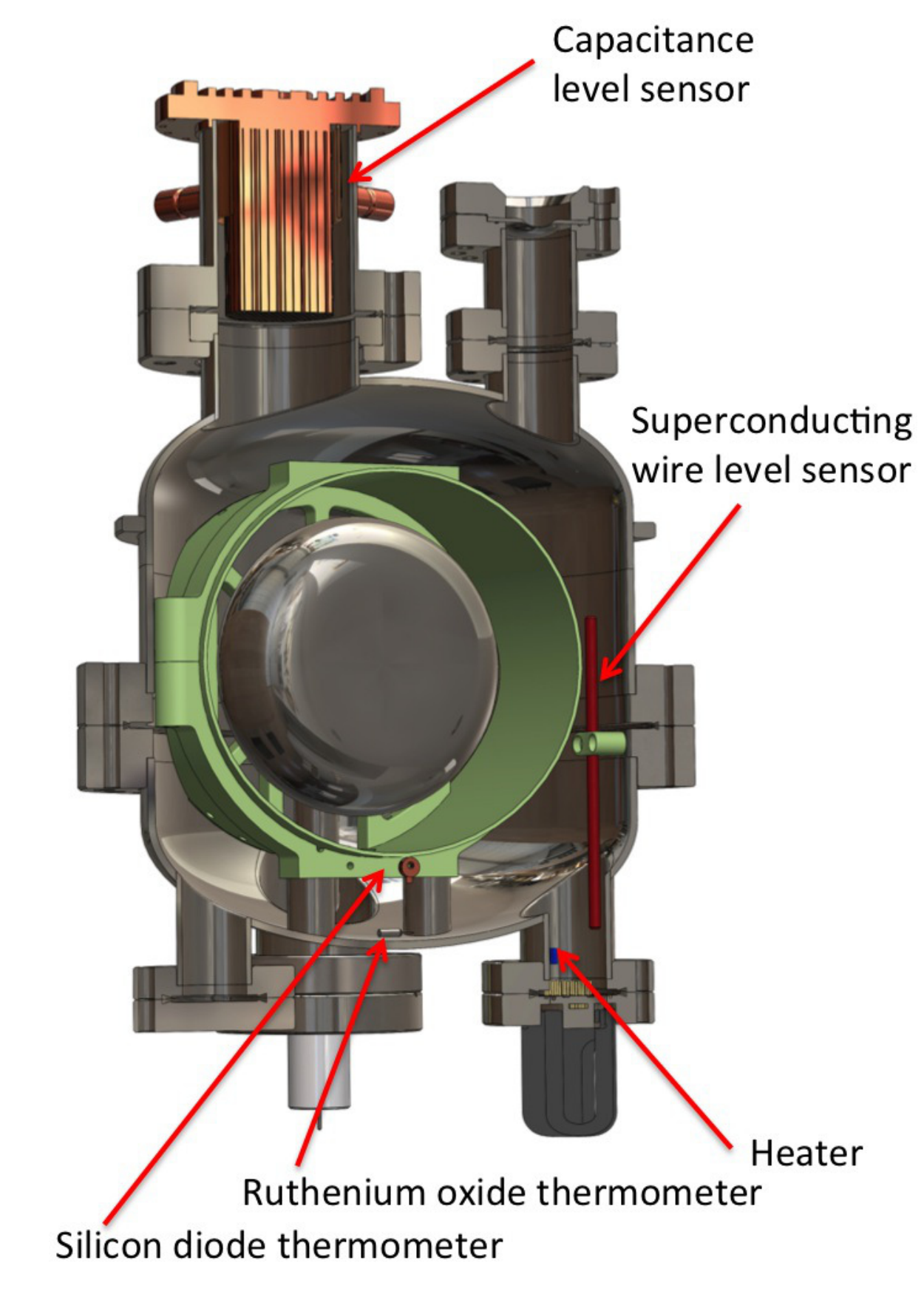}
\caption{The design of the CV, showing the locations of a capacitance
  level sensor, superconducting wire level sensor, and two
  thermometers .\label{fig:CV}}
\end{figure}

A schematic showing the design of the CV, along with the location of
some of the sensors, is shown in Fig.~\ref{fig:CV}.

\section{Design of HV components}
\label{sec:HVDesign}
As mentioned earlier, there are two HV feed lines, each providing HV
up to $\pm$50~kV to the corresponding electrode. For each feed line,
there are four HV feedthroughs. They are:
\begin{itemize}
\item FT1: feedthrough to go from atmosphere to the insulating vacuum,
  located at room temperature.
\item FT2: feedthrough to go through the 77~K shield. This feedthrough does
  not have to be leak tight. Its purpose is to thermally anchor the HV
  feed line to the 77~K shield.
\item FT3: feedthrough to go through the IVC/4~K shield. This
  feedthrough needs to be vacuum tight as the IVC needs to hold the
  exchange gas.
\item FT4: feedthrough to go from the IVC vacuum to LHe inside the
  CV. This feedthrough needs to be superfluid tight. 
\end{itemize}
There are also HV lines that connect between these feedthroughs. 

Note that it is very important that heat leaks to the HV electrodes be
minimized as they can cause vapor bubbles to be created on the surface
of the electrodes, which in turn can initiate electrical breakdown,
potentially leading to erroneous results, as mentioned earlier. The
heat leaks to the HV electrodes were minimized by thermally anchoring
the HV feed lines at each heat shield and by choosing a design and
materials for the HV feed line that minimizes conductive heat leak.

\subsection{HV feedthroughs}
For FT1, we chose CeramTec\cite{CERAMTEC} 100~kV feedthroughs (part
no. 6722-01-CF). The reason for choosing 100~kV feedthroughs instead
of 50~kV ones was the desire to be able to supply as high a potential
as possible to each electrode. While the initial design goal was to
achieve up to $\pm$50~kV for the HV feed line for each electrode, the
plan was to make improvements where possible.

\begin{figure}
\centering
\includegraphics[width=3.in]{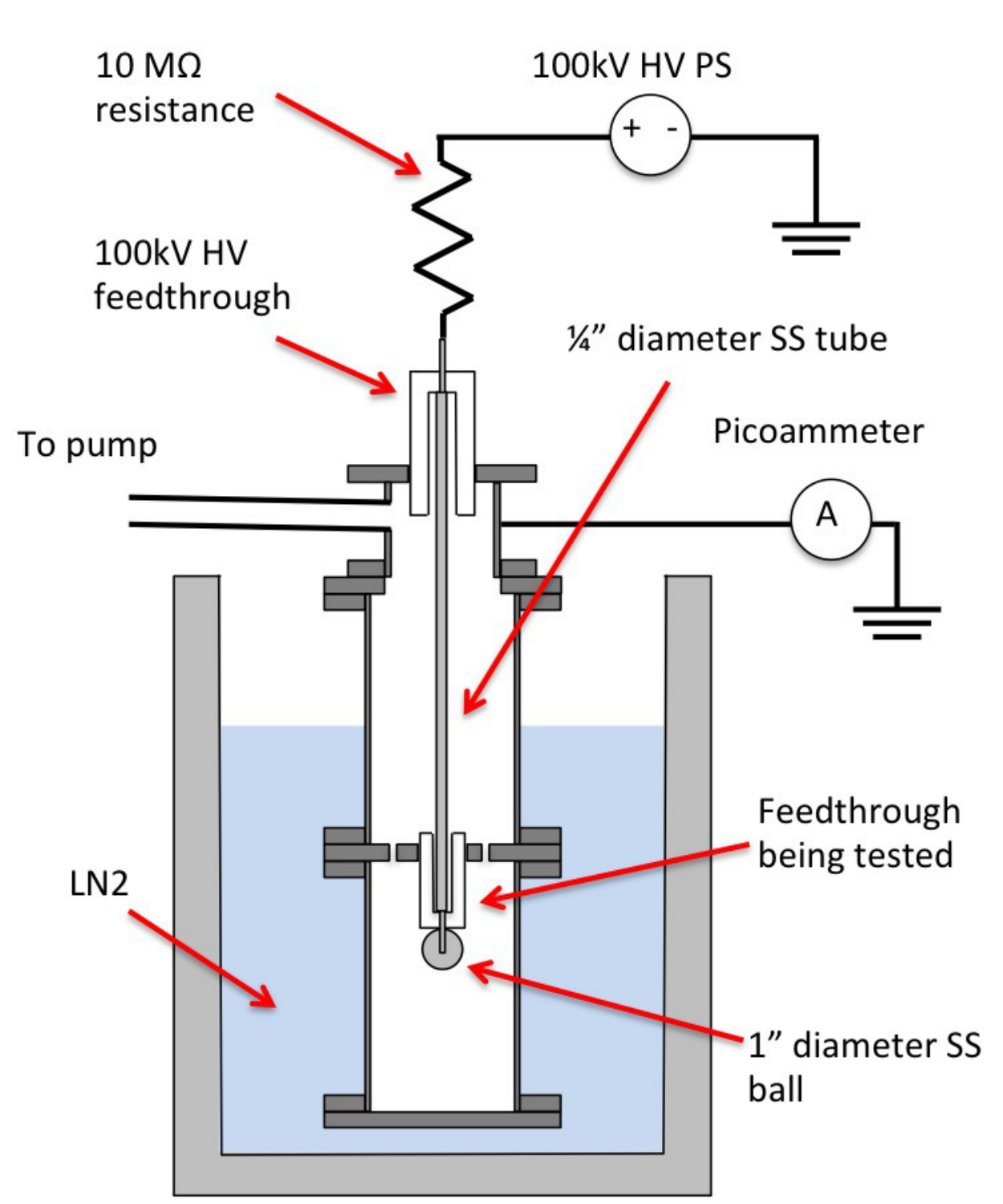}
\caption{A schematic of the apparatus used for the measurement of
  surface leakage currents of CeramTec 50~kV feedthroughs (part
  no. 21183-01-W) as a function of the voltage applied to the central
  conductor at 300~K and 77~K. The tested feedthrough was kept under
  vacuum. \label{fig:FTtest_apparatus}}
\end{figure}

For FT2, FT3, and FT4, due to spatial constraints, we chose CeramTec
50~kV feedthroughs (part no. 21183-01-W). This model is rated for use
at 4~K. Prior to installing them in the MSHV apparatus, we performed
measurements to characterize their performance for the following
purposes:
\begin{itemize}
\item To determine the real performance limitation of these
  feedthroughs, as we are interested in applying as high a voltage as
  possible.
\item To study how the performance of the feedthrough depends on
  the temperature, as some of the processes responsible for surface
  leakage currents and surface flashover may be suppressed at lower
  temperature.
\item To study the dependence of the feedthrough performance on the
  method used to clean the ceramic. 
\end{itemize}
We measured the (surface) leakage currents as a function of the
voltage applied to the central conductor at 300~K and
77~K. Figure~\ref{fig:FTtest_apparatus} shows a schematic of the
apparatus used for this test.

\begin{figure}
\centering
\includegraphics[width=3.in]{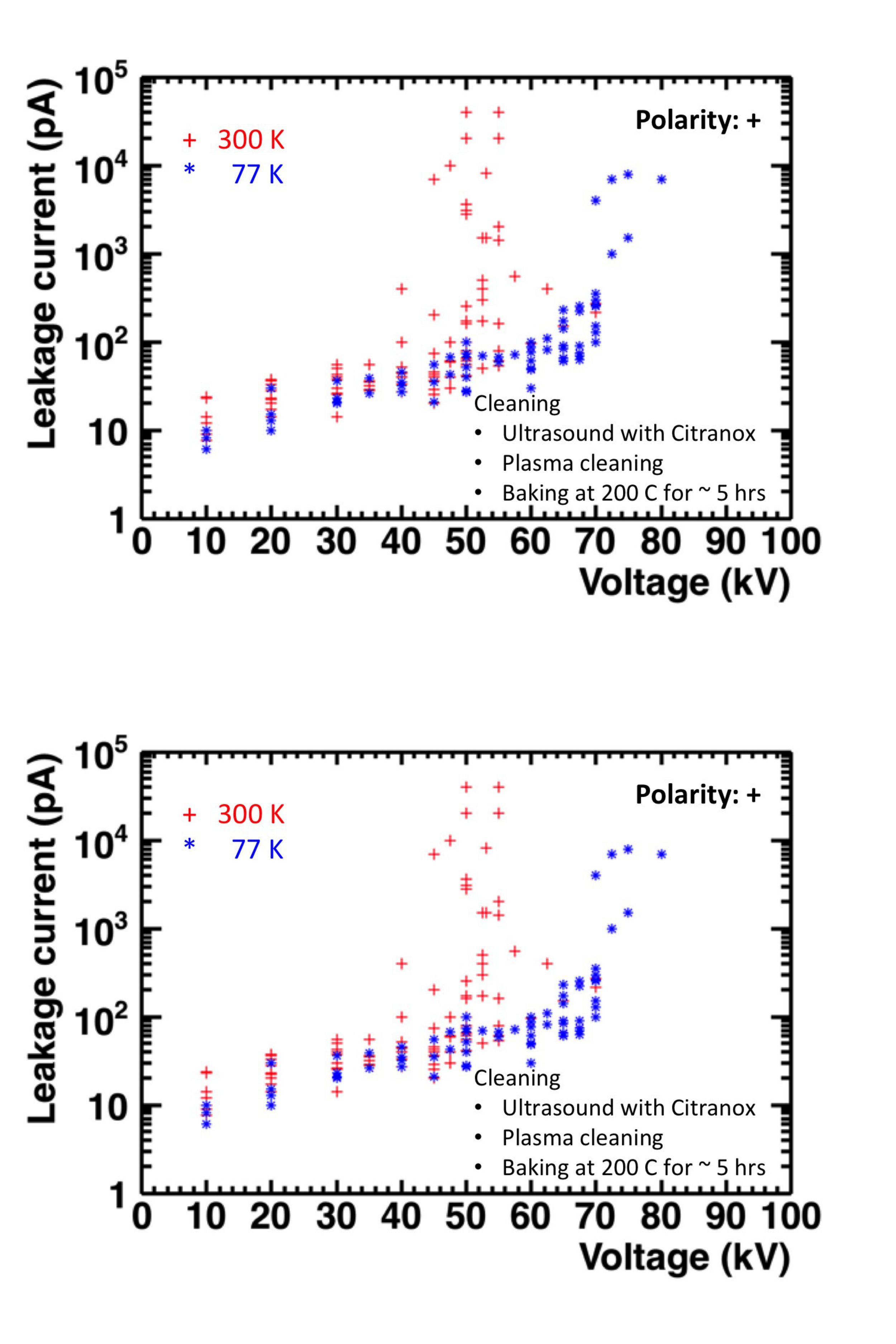}
\caption{Leakage currents measured for CeramTec 50~kV feedthroughs
  (part no. 21183-01-W) as a function of the voltage applied to the
  central conductor at 300~K and 77~K. Top: the feedthrough was first
  cleaned in an ultrasonic bath with Citranox, then was cleaned with
  plasma cleaner, and was baked at 200~C for 5 hours. Bottom: the
  feedthrough was cleaned with a mixture of alumina powder and ethanol
  (purity $>99.5$\%), then was rinsed with ethanol (purity $>99.5$\%),
  and then was baked at 200~C for 5 hours.\label{fig:FTtest}}
\end{figure}

Figure~\ref{fig:FTtest} shows results obtained at 300~K and 77~K for
the following two methods of cleaning:
\begin{enumerate}
\item The feedthrough was first cleaned in an ultrasonic bath with
  Citranox, then cleaned with a plasma cleaner, and baked at
  200$\degree$C for 5 hours.
\item The feedthrough was cleaned with a mixture of alumina powder and
  ethanol (purity $>99.5$\%), rinsed with ethanol (purity
  $>99.5$\%), and baked at 200~C for 5 hours.
\end{enumerate}
The results indicate that the leakage current is smaller for 77~K than
for 300~K. Also the performance of the feedthrough depends on the
cleaning method employed. When the feedthrough was cleaned using the
second method described above, it was possible to apply up to 90 kV
with a leakage current less than 1~nA. All the feedthroughs used for
FT2, FT3, and FT4 were prepared in this way prior to installation in 
the MSHV system.

\subsection{HV interconnects}
\label{sec:HVlines}
The HV lines connecting between the feedthroughs inside the cryostat need
to meet the following requirements:
\begin{itemize}
\item The conductive heat leak through the HV lines should be
  minimized. 
\item The HV lines need to have some amount of flexibility for
  installation and also for accommodating the thermal contraction of
  different parts of the system. 
\item The initial goal is to have a HV performance of $|V|\geq 50$~kV.
\end{itemize}

To meet these requirements, we settled on a design in which the HV
lines are made of thin wall stainless steel tubing (6.35~mm diameter,
0.04~mm thick wall, 304~SS tubing), and formed bellows where
flexibility is needed. Furthermore, a smooth metal sheath was slipped
over each bellows section so as to avoid the ridges of the bellows
causing high electric fields. The validity of such a design was
confirmed using finite element electrostatic calculation, performed
using COMSOL.\cite{COMSOL} Additionally, the surface of the conductor
was mechanically polished and was covered with Stycast 2850 FT Blue.

\begin{figure}
\centering
\includegraphics[width=3.in]{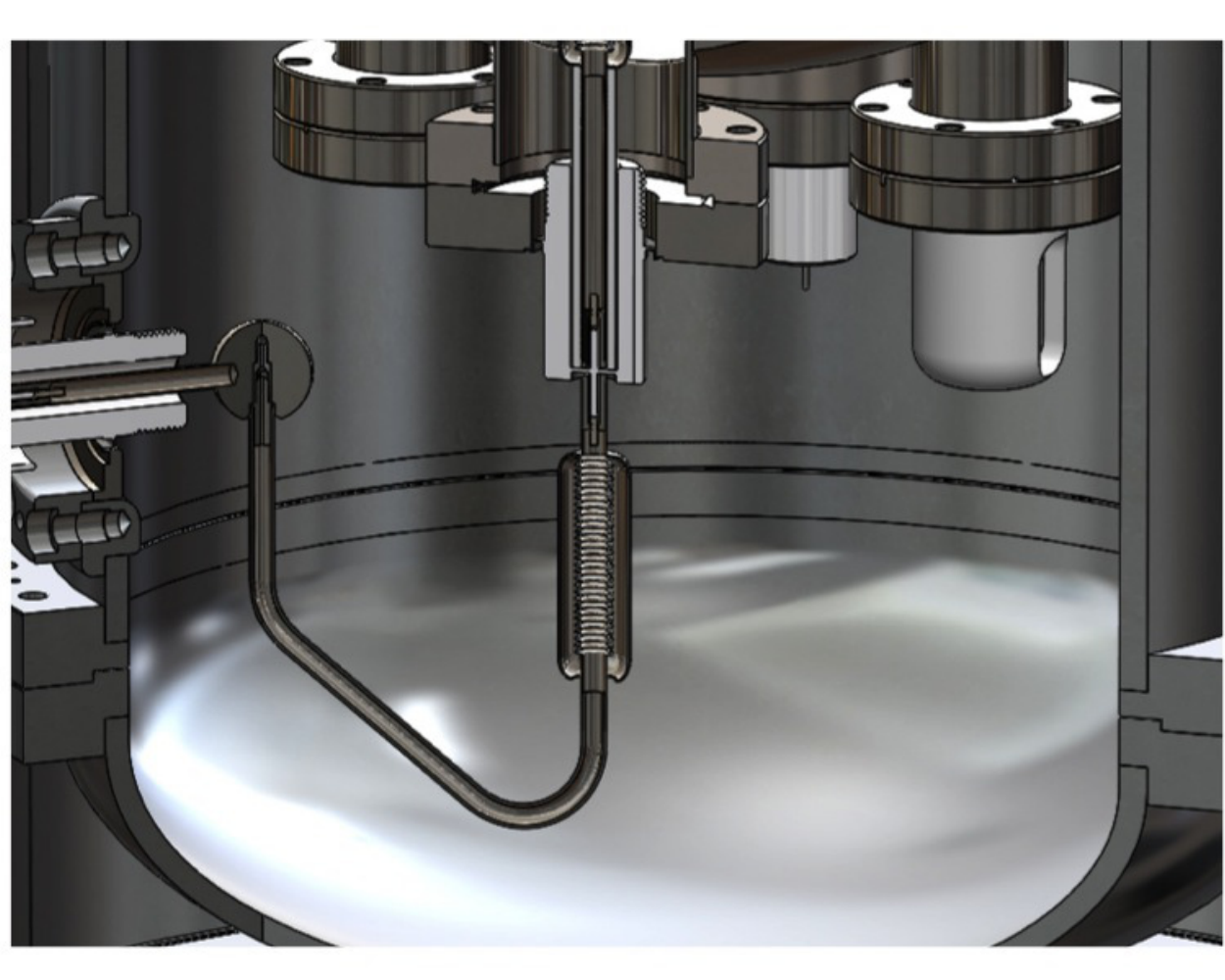}
\caption{The design of the HV interconnecting line between FT3 and
  FT4. \label{fig:HVline}}
\end{figure}

\begin{figure}
\centering
\includegraphics[width=3.in]{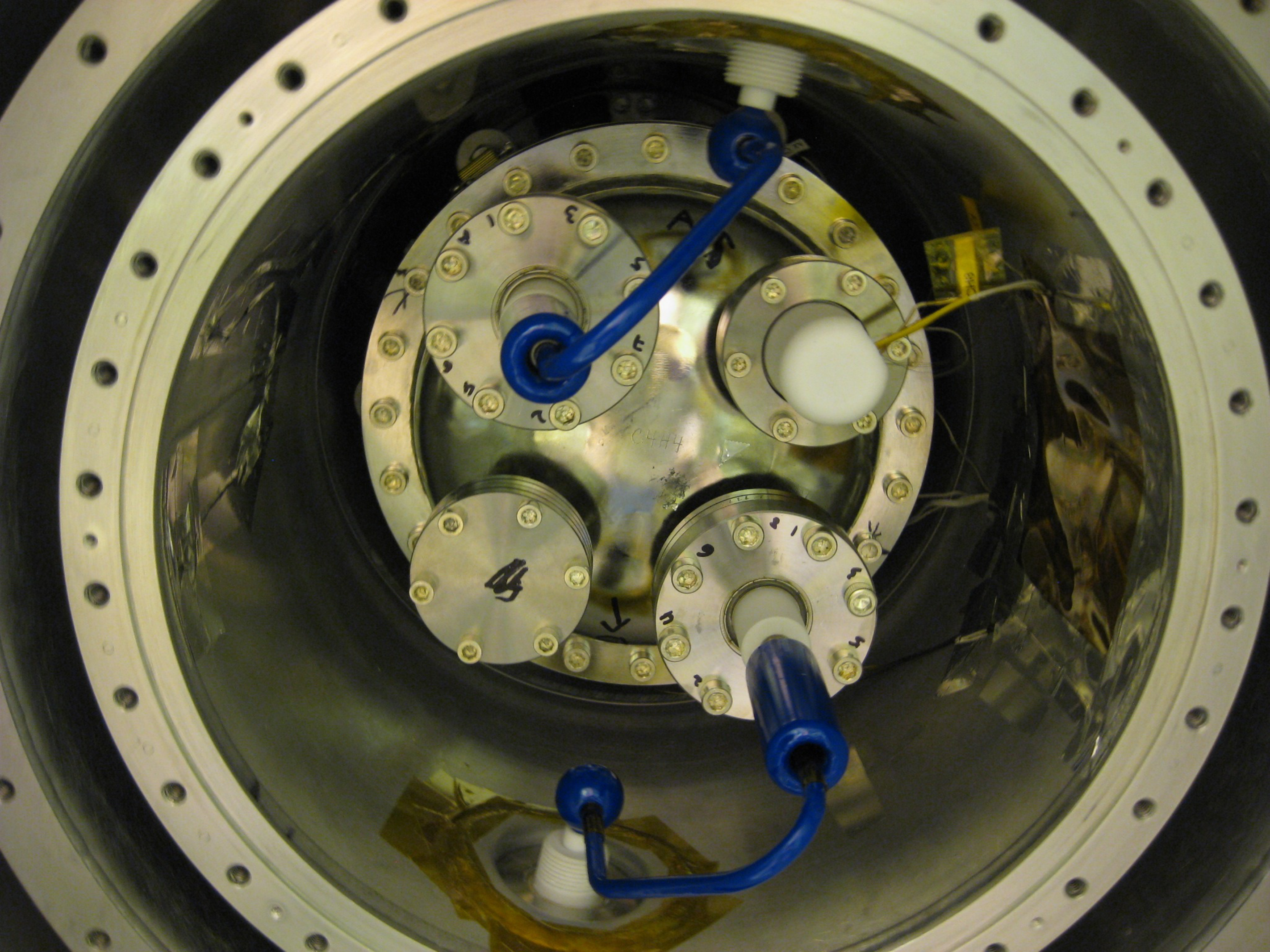}
\caption{Photograph of the bottom of the CV, which shows the HV
  lines from FT3 to FT4.\label{fig:HVinterconnects}}
\end{figure}

The design of the HV interconnecting line between FT3 and FT4 is shown
in Fig.~\ref{fig:HVline}. Figure~\ref{fig:HVinterconnects} shows a
photograph of the bottom of the CV, which shows the HV lines from FT3
to FT4.
The HV lines from FT4 to the electrodes were designed in a similar
manner to the HV lines between the feedthroughs.
 
\subsection{Power supply and HV resistor}
We used two Model AF-100R0.1 100~kV DC HV power supplies from
Matsusada Precision Inc. HV was supplied to each HV feed line through
a 1~G$\Omega$ HV resistor in order to limit the energy released inside
the CV when an electrical breakdown occurs. The energy stored in the
system is dominated by that stored in the room temperature HV cable
connecting the power supply to the MSHV system, which has a
capacitance of $\sim$300~pF. Therefore, the HV resistance was placed
between FT1 and the room temperature HV cable.

\subsection{Rogowski electrodes}
\label{sec:rogowski}
The goal of the initial HV test was to establish the electric field
strength that is possible with an electrode material known to be
capable of good performance. Therefore, the initial electrodes were
made of electropolished stainless steel. Stainless steel is one of the
materials that are known to give higher breakdown fields than others.

\begin{figure}
\centering \includegraphics[width=3.5in]{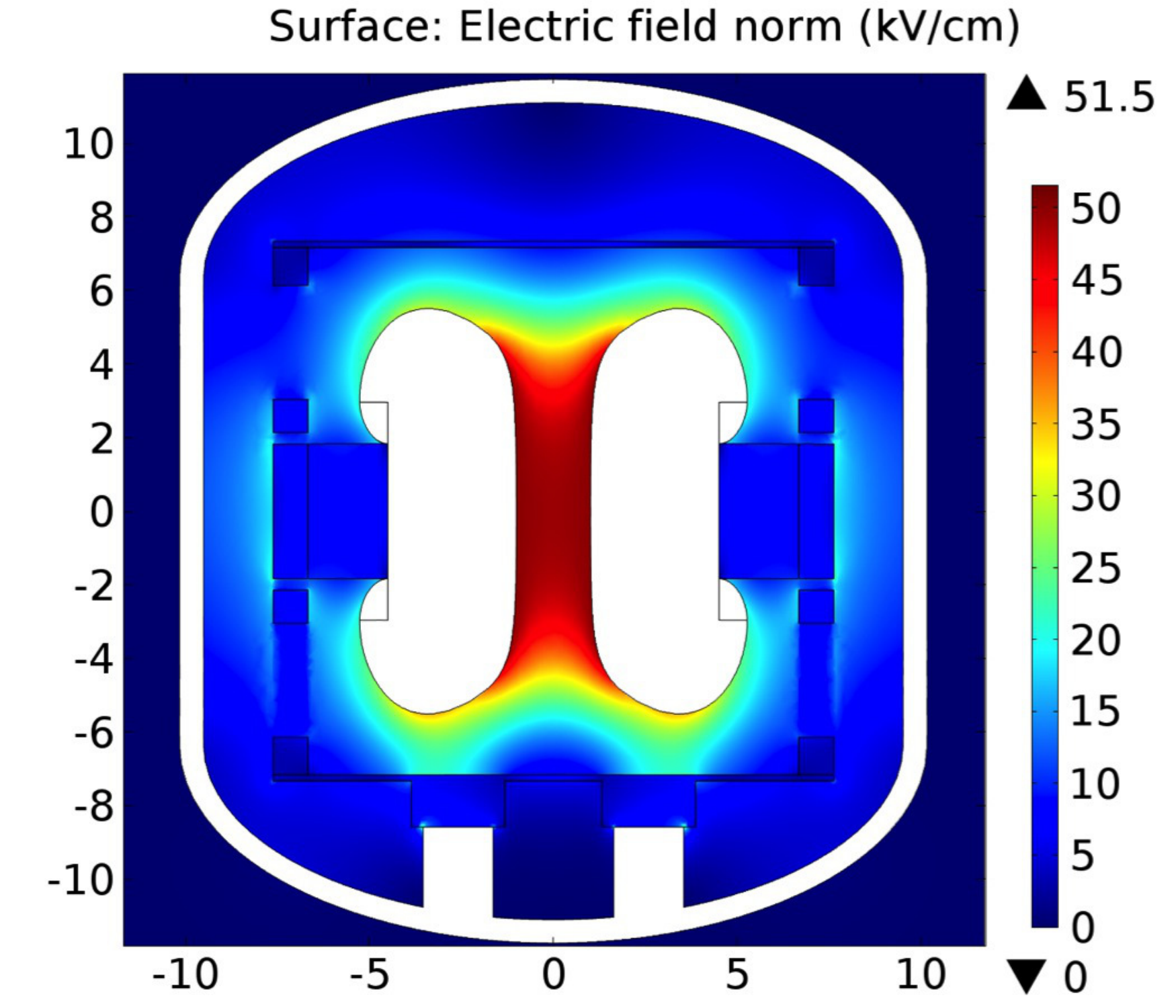}
\caption{Result of a finite element electrostatic calculation,
  performed using COMSOL,\cite{COMSOL} of the Rogowski electrodes
  mounted inside the CV. For this calculation, the gap betweeen the
  two electrodes was set to 2~cm and the two electrodes were held at
  +50~kV and $-50$~kV.\label{fig:electrodes}}
\end{figure}

For the shape of the electrodes, we chose the so-called Rogowski
profile,\cite{COB58} which provides a uniform electric field in the
gap and ensures that the gap has the highest field in the
system. Figure~\ref{fig:electrodes} shows a result of a finite element
electrostatic calculation, performed using COMSOL,\cite{COMSOL} of the
Rogowski electrodes that we used. We also performed finite element
electrostatic calculations for other known uniform field electrode
(UFE) shapes, such as that by Chang\cite{CHA73} and that by
Ernst.\cite{ERN83} We did not see a significant difference among these
UFE shapes in terms of the uniformity of the electric field in the gap
and of the area over which the field is uniform. For the chosen
geometry, the field is uniform to 2\% over $\sim$64~cm$^2$.

We decided to electropolish the electrodes based on the following
considerations. A protrusion on the electrode surface would cause a
spot with higher electric field (a ``hot spot''). The strength of the
field on such a hot spot does not depend on the size of the protrusion
but on its shape, as long as the size of the protrusion is much
smaller than the distance between two electrodes. For example, if a
half sphere is embedded on a surface of a flat electrode that forms a
parallel plate capacitor and the field between the two plates is $E_0$
in the absence of the half sphere, then the field on the top of the
half sphere is $3E_0$, regardless of the size of the half sphere, as
long as the radius $r$ of the sphere is much smaller than the distance
$d$ between the two plates.\cite{LEN81} The lack of dependence of the
field strength on the size of the sphere can be easily understood by
realizing that there is no length scale in the problem when $r<<d$. It
follows that the amount of the field strength enhancement due to
features on the electrode surface depends on the shape of the features
but not on the size. Surface smoothness is typically expressed in
terms of the size of the features on the surface and mechanical
polishing processes reduce the size of the features. However, it is
not the relevant quantity in this case. Electropolishing on the other
hand is expected to make the shape of the features smoother. For this
reason, we decided to electropolish the first set of electrodes,
although the difference between electropolished surfaces and machine
polished surfaces needs to be ultimately studied experimentally.

\begin{figure}
\centering
\includegraphics[width=3.in]{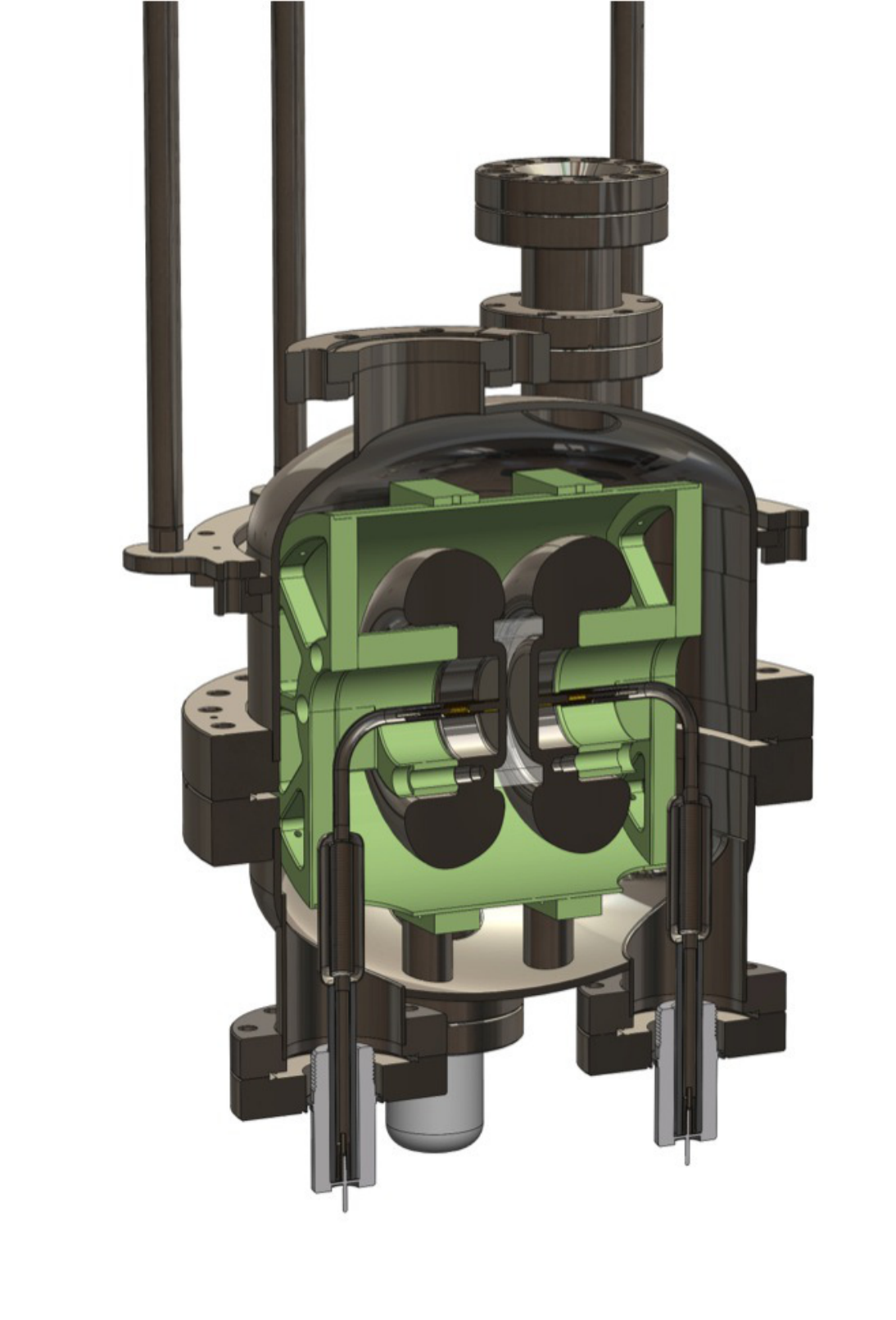}
\caption{An image showing how the electrodes were supported inside the
  CV. The grooved electrodes with a PMMA ring are
  shown. \label{fig:CV_and_electrodes}}
\end{figure}

Figure~\ref{fig:CV_and_electrodes} shows how the electrodes were
mounted inside the CV. The electrodes were supported by a ``cage''
made of fiberglass-reinforced plastic (FRP), in this case, G-10. A
spacer ring made of FRP was bolted onto the back of each electrode
with a set of unfilled Torlon bolts. The spacers were then mounted
onto the end caps of the FRP cage using Torlon bolts.  With such a
supporting structure, the electrodes were expected to remain parallel
to each other to $\sim$0.1$\degree$ when cooled. With the aid of
COMSOL-based electrostatic calculations, we determined that with the
expected tolerance within which the two electrodes can be located, the
effect of possible misalignment (both angular and translational) of
the electrodes on the field uniformity and the area over which the
field is uniform would be insignificant. The effect of the supporting
structure on the electric field distribution was also evaluated as
shown in Fig.~\ref{fig:electrodes}. The HV lines were included
in the electrostatic modeling to see the strength of the field on the
surface.

\subsection{Grooved electrodes}
\label{sec:grooved}
The second test we performed was to study the effect of having a
dielectric object sandwiched between two electrodes. As indicated
earlier, it is well known that when a dielectric object is inserted
between two electrodes, the holdoff voltage is reduced, which is
thought to be due to field emission at the cathode-insulator junction
and emitted electrons running along the insulator surface.\cite{KOF60}
A traditional solution to this problem is to embed the end of the
insulator into a groove.\cite{KOF60,GOL86} The idea is to ``hide'' the
edges of the insulator in a low field region to avoid the enhancement
of the field at the cathode-insulator junction due to the higher
dielectric constant of the dielectric. Designs in which the insulator
surface is at 45$\degree$ with respect to the flat electrode surface
are also used, in particular for pulsed high-power high-voltage
applications.\cite{STY05} However the latter method is not suitable 
for our application.

\begin{figure}
\centering
\includegraphics[width=3.in]{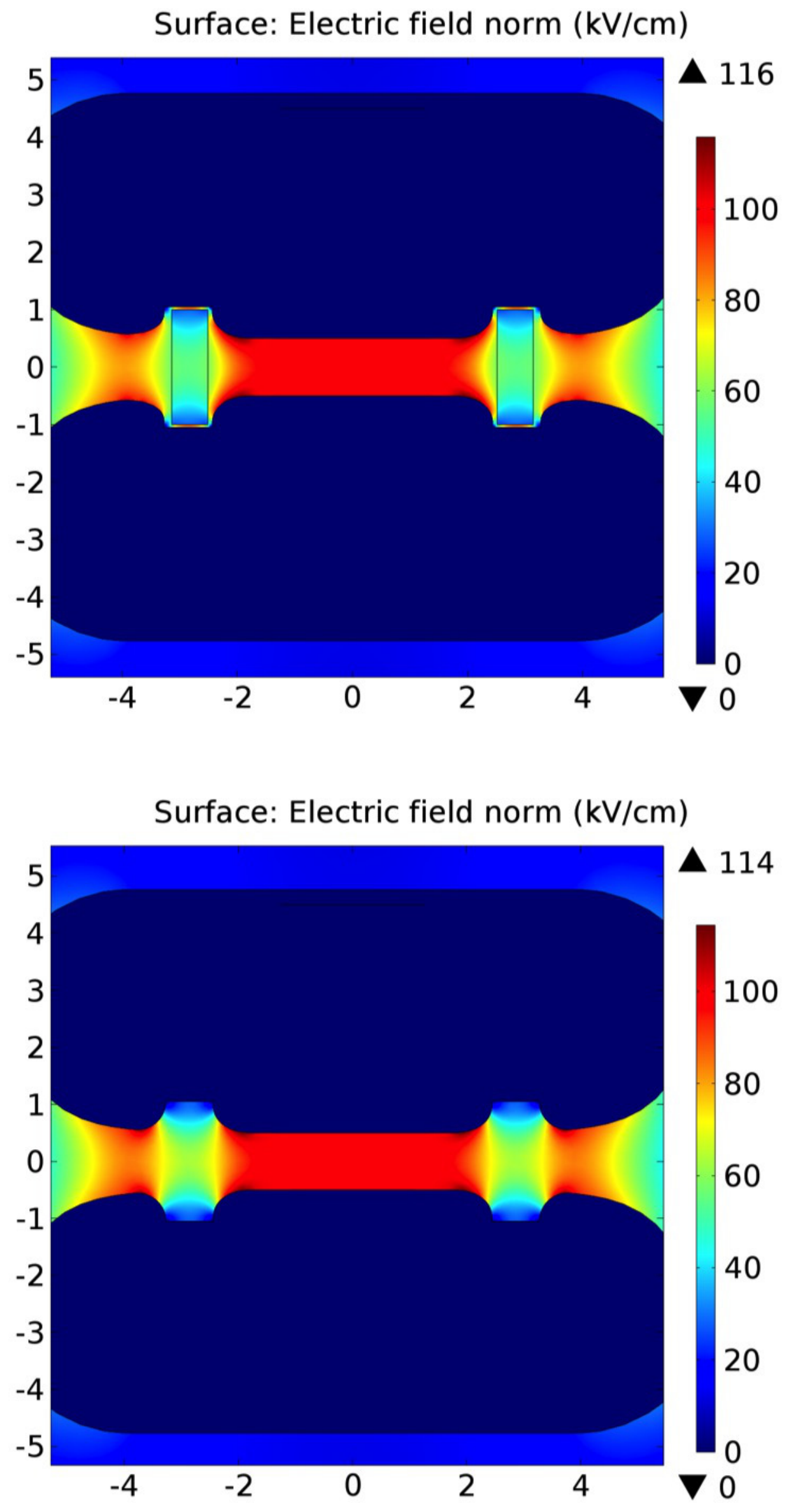}
\caption{Electrostatic models of the grooved electrodes. Top: with the
  PMMA insert. Bottom: without the PMMA
  insert.For these calculations, the gap betweeen the
  two electrodes was set to 1~cm and the two electrodes were held at
  +50~kV and $-50$~kV.\label{fig:grooved_electrodes}}
\end{figure}

\begin{figure}
\centering
\includegraphics[width=3.in]{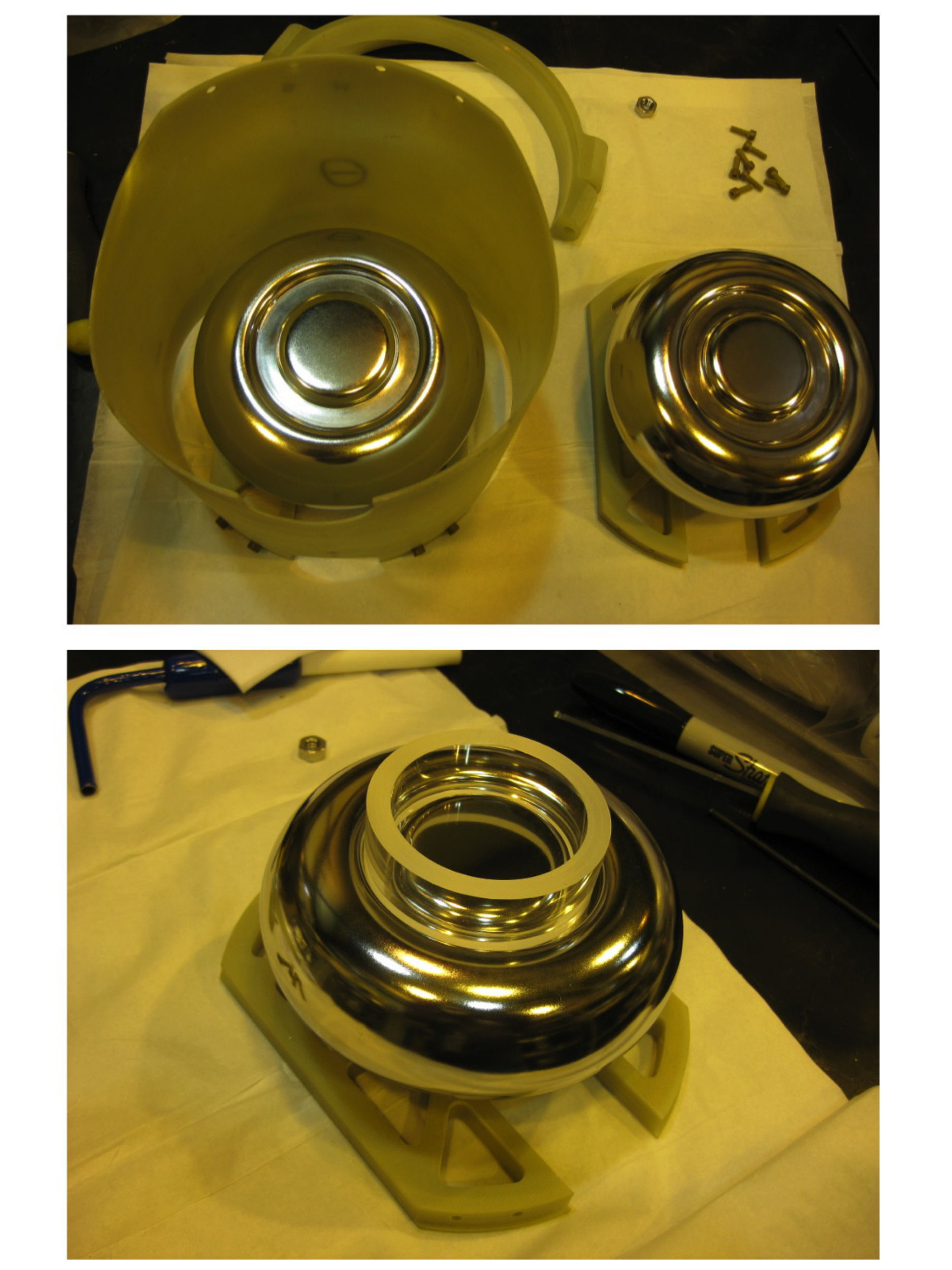}
\caption{Photographs of the grooved electrodes and the PMMA
  insert\label{fig:grooved_electrodes_photos}}
\end{figure}

We used a PMMA cylinder 6.35~cm in OD and 5.08~cm in ID as our
dielectric insert. We designed a set of electrodes with a groove with
the help of finite element electrostatic analysis using
COMSOL.\cite{COMSOL} Figure~\ref{fig:grooved_electrodes} shows results
from such electrostatic calculations. The shape of the groove was
adjusted so that the highest field in the system, which occurs on the
rounded corner of the groove, is less than $\sim$1.15 times the
uniform field in the middle of the electrodes with and without the
dielectric insert. We started with a groove 5.5~mm deep and 8.4~mm
wide (2~mm larger than the insulator's wall thickness), and adjusted
the shape of the edge of the groove by trial and error using COMSOL's
chamfer and fillet tools in order to achieve the necessary field
profile. In doing so, the following competing factors were taken into
consideration:
\begin{enumerate}
\item If the rounding of the edge of the groove has too small a
  radius, then there will be a high field region between the rounded
  edge of the groove and the side wall of the PMMA insert. Whereas the
  surface of the electrode, including the rounded edge of the groove,
  forms an equipotential surface, there is a potential gradient along
  the surface of the PMMA insert. Therefore, the distance between the
  electrode surface and the PMMA insert surface being too close at the
  location of the rounded edge of the groove causes the field there to
  be high.
\item If the rounding of the edge of the groove has
too large a radius, then the field at the bottom of the groove does
not become sufficiently low. 
\end{enumerate}
Figure~\ref{fig:grooved_electrodes_photos} shows photographs of the
grooved electrodes and the PMMA insert.

\section{Cryogenic performance and operational experience}
At the time of this writing, the system has been operated 18 times
successfully. In this section, we describe the cryogenic performance
of the system and our operational experience with the system. As
mentioned earlier, one of the design goals was to achieve a turnaround
time of 2~weeks. We have demonstrated that indeed turning around the
system in two weeks was possible, including the time to replace the
contents of the CV.

\label{sec:CryoPerformance}
\subsection{Cooldown and temperature control}
The cooldown of the system starts with precooling the LHe bath with
liquid nitrogen (LN$_2$) as is done as a standard practice for many
LHe cryostats. In doing so, the IVC is filled with 1~atm N$_2$ gas as
the exchange gas to cool the 1~K pot, the $^3$He pot, and the CV. Once
the system reaches LN$_2$ temperature, which takes $\sim$12~hours
after the LHe bath is filled with LN$_2$, the remaining LN$_2$ is
removed from the LHe bath and the N$_2$ exchange gas is pumped out of
the IVC volume. 

The system is further cooled by transferring LHe (initially cold
helium gas) to the LHe bath. During this time, the CV is cooled by
opening both MV1 and MV2 and flowing LHe (initially cold
helium gas) from the LHe bath through the CV. The 1~K pot is also
pumped on at the same time. As the CV is cooled sufficiently, the CV
gets filled with LHe. It takes approximately 6~hours for the CV to
become completely filled from the time when LHe starts to accumulate
at the bottom of the CV. Once the CV becomes completely filled with
LHe, MV1 is closed. When the CV becomes full, the temperature of
the CV is $\sim$1.8~K. We also tried filling the IVC with helium
gas as the exchange gas to cool the CV, instead of flowing cold helium
gas through the CV. We found that flowing cold helium gas and LHe
through the CV is a more efficient way of cooling it than using an
exchange gas, because the use of exchange gas required spending time
pumping out the exchange gas. 

The $^3$He refrigerator is then turned on and the $^3$He circulation
starts. This further cools the CV. It takes approximately 4 hours to
reach 0.4~K. Overall, it takes $\sim$2.5~days to go from the CV being
empty at 300~K to the CV being filled with 0.4~K LHe.

The temperature of the CV can easily be controlled either by
adjusting the $^3$He circulation or by using heaters that are mounted
inside the CV (directly in contact with LHe) and on the outside wall
of the CV.

\subsection{Pressure control and monitoring}
As mentioned earlier, leaving MV2 open keeps the pressure of the LHe
inside the CV at the pressure of the LHe inside the 4~K LHe bath,
which is typically 600~torr. Closing MV2 and pumping on the fill line
to remove LHe from the fill line allows the pressure in the CV to be
reduced. The pressure can be increased by stopping pumping on the fill
line and opening MV2 momentarily. The pressure in the CV will be the
SVP of LHe at the location of the temperature of the liquid surface,
which is somewhere on the fill line, along which there is a
temperature gradient. The pressure can be changed and controlled with
ease between $\sim$1~torr and $\sim$600~torr with the method described
above. Going below $\sim$1~torr cannot be achieved efficiently by
simply pumping on the fill line, because the vapor pressure of helium
becomes smaller and smaller. However, lower pressures can be achieved
by removing LHe from the fill line by first warming up the CV while
pumping on the fill line and cooling it again. Warming up the CV to
higher temperatures makes it easy to remove LHe from the fill line
because of the higher vapor pressure. Note that the hydrostatic
pressure of LHe is $\sim$0.1~torr/cm. Therefore even if the LHe
surface is below the top of the CV, in which case the lowest pressure
inside the CV is $\sim$10$^{-6}$~torr, the pressure at the location of
the electrodes is $\sim$1~torr.

The pressure inside the CV can be monitored by the external pressure
gauge (Mensor model 2101), as indicated in Fig.~\ref{fig:plumbing}. In
addition, the pressure change inside the CV can be monitored using the
capacitance LHe level sensor mounted around the heat exchanger. The
change in the dielectric constant due to the change in pressure is
large enough to make the capacitance level sensor sensitive to the
pressure change. The dielectric constant $\epsilon$ is related to the
density $\rho$ via the Clausius-Mossotti relation
\begin{equation}
\label{eq:C-M}
\frac{\epsilon - 1}{\epsilon + 2} =
\frac{4\pi\alpha_M \rho}{3M},
\end{equation}
or solving for $\epsilon$,
\begin{equation}
\label{eq:C-M2}
\epsilon = \frac{1+2\eta}{1-\eta},\;\;\;\eta=\frac{4\pi\alpha_M\rho}{3M},
\end{equation}
where $M$ is the molecular weight and $\alpha_M$ is the molar
polarizability. The density $\rho$ depends on the pressure and
temperature. $\alpha_M$ has a very weak pressure dependence. Since
$\eta<<1$ and the density increases approximately linearly with the
pressure in the pressure range we are interested in
($p\lesssim1$~atm), it is expected that the value of $\epsilon$ and
therefore the capacitance of the LHe level sensor depends
approximately linearly on the pressure.

\begin{figure}
\centering
\includegraphics[width=3.5in]{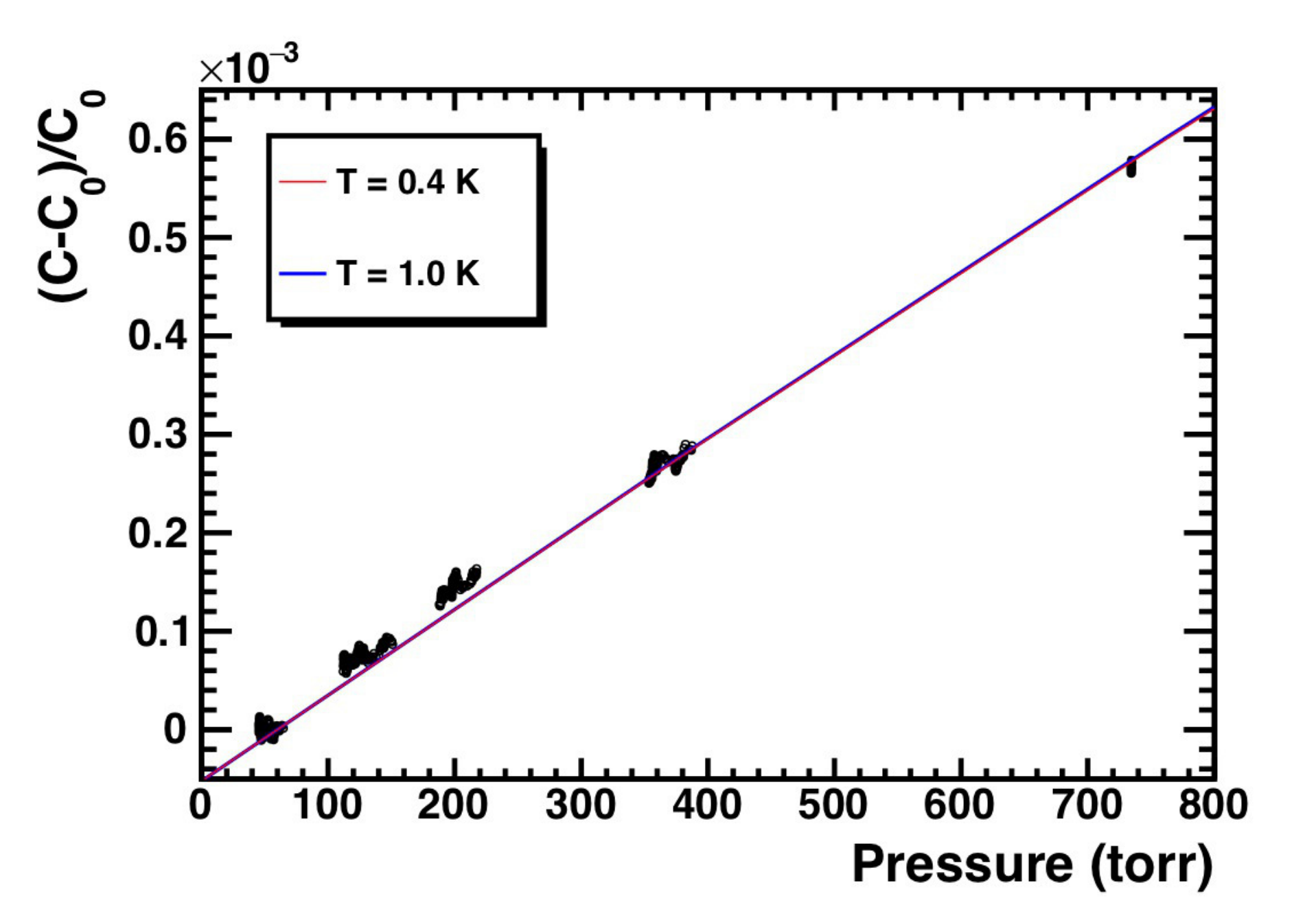}
\caption{The measured capacitance output of the LHe level sensor
  plotted as a function of the pressure of LHe inside the CV as
  measured by an external pressure gauge. C$_0$ refers to the
  capacitance measured at the reference pressure, which in this case
  was 60~torr. Also shown is the expected change in capacitance as a
  function of pressure, calcualted from Eq.~(\ref{eq:C-M2}) using molar
  volume data from Ref.~\onlinecite{TAN00} and assuming a constant
  $\alpha_M$ from Ref.~\onlinecite{HAR70} for two different
  temperatures, illustrating insensitivity to temperature in this
  range.\label{fig:p_vs_c}}
\end{figure}

In Fig.~\ref{fig:p_vs_c}, the measured change in capacitance is
plotted as a function of the pressure as measured by the external
pressure gauge and is compared to what is expected from
Eq.~(\ref{eq:C-M2}) with molar volume data from
Ref.~\onlinecite{TAN00} and assuming a constant $\alpha_M$ from
Ref.~\onlinecite{HAR70}. The relative change in capacitance is
plotted, rather than the value of the dielectric constant, because we
do not have sufficiently accurate information on the geometry of the
capacitor to extract the value of the dielectric constant from the
measured capacitance. However, these results demonstrate that the
capacitance level LHe meter serves as an internal pressure gauge with
precision sufficient for the our purpose.

\subsection{Removal of LHe and warming up}
When warming up the system, LHe from the CV can be removed by
opening MV1, pumping on the CV, and turning on the heater in the
CV, after stopping the $^3$He circulation. With a heater power of
$\sim$0.6~W, it takes $\sim$8 hours to remove LHe from the
CV. By spoiling the insulation vacuum with clean neon gas after all
the cryogen is removed from the system, the system can be warmed up to
room temperature in $\sim$2~days. 

\section{HV performance, operational experience, and initial results}
\label{sec:HVPerformance}
\subsection{Holdoff voltage}
For the initial HV tests, we operated the HV power supplies
manually. Typically we increased the HV at a rate of 0.5~kV/s or less
pausing every 5 or 10 kV. The current monitor output of the power supplies,
which gives a 10~V output for a 0.1~mA current, was connected to a
digital oscilloscope for monitoring and recording. Electrical breakdown
events with currents of $\sim$0.1~$\mu$A were easily detected. Initial
tests indicated that the performance of the HV feedlines was limited
to $\sim \pm40-50$~kV. Applying a higher voltage causes breakdown from
the HV feedline to the ground. Likely locations for this are inside
the CV. 

For neither of the two electrode configurations discussed in
Sec.~\ref{sec:rogowski} and Sec.~\ref{sec:grooved} was electrical
breakdown between the two electrodes observed. In both cases, the
applicable HV was limited by the performance of the HV feedlines. For
the case of the electropolished stainless steel Rogowski electrodes,
we were able to stably apply up to 105~kV across the 1~cm gap for a
wide range of pressures and temperatures. For the electropolished
stainless steel electrodes with a groove, we were able to stably apply
up to 80~kV across the 1~cm gap for a wide range of pressures and
temperatures, with and without the PMMA insert. Note that the HV
feedlines inside the CV were different between these two cases,
because of the slightly different design of the back of the
electrodes. These initial results are summarized in
Table~\ref{tab:HVresults}.

\begin{table*}
\caption{Summary of holdoff voltage results. Note that in all cases
  except noted, the gap between the electrodes was 1 cm and the
  highest achievable stable field was limited by the performance of
  the HV feedlines.\label{tab:HVresults}}
\begin{tabular}{lccc}\\ \hline\hline
Configuration &  Temperature & Pressure & Highest achievable stable field 
\\ \hline
SS Rogowski electrodes & 2.5-2.8~K & 72 torr, 90 torr, 785 torr & 90~kV/cm \\
SS Rogowski electrodes & 1.1~K & 13 torr & 90~kV/cm \\
SS Rogowski electrodes & 0.42~K & SVP, 664 torr& 105~kV/cm  \\ \hline
SS electrodes with a groove without the PMMA insert & 2.7~K & 620 torr
& 70~kV/cm  \\
SS electrodes with a groove without the PMMA insert & 0.5~K
& 1.0~torr, 620 torr& 80~kV/cm \\ \hline
SS electrodes with a groove with the PMMA insert &
2.1~K & 614~torr & 50~kV/cm$^*$ \\ 
SS electrodes with a groove with the PMMA insert &
0.5~K & SVP, 2.7~torr, 574~torr & 80~kV/cm \\ \hline\hline
\end{tabular}
$^*$Higher voltages were not tried. 
\end{table*}

\subsection{Leakage currents}
As discussed in the introduction, in the SNS nEDM experiment the
leakage currents along the cell walls need to be minimized.  Processes
responsible for leakage currents may be suppressed at cryogenic
temperatures.  It is therefore important to study the leakage currents
when a dielectric object is sandwiched between two electrodes in the
MSHV system.

As a first step, we measured the current flowing between the two
electropolished stainless steel Rogowski electrodes ({\it i.e.} no
dielectric objects between two electrodes). The measurement was done
by applying HV to one electrode and measuring the current coming out
of the other electrode using a picoammeter. In this configuration, the
maximum voltage difference between the two electrodes was half that
possible when both electrodes are connected to the respective HV power
supply. The reason for doing this is that the leakage current through
the surface of the feedthrough insulators and that through the
insulator of the room temperature HV cables are much larger than the
current expected to flow between the two electrodes.

The measured current was less than 1~pA with 45~kV applied to one of
the electrodes. The measurement was limited by the sensitivity of the
current measuring device. This gives a lower bound on the effective
volume resistivity of LHe of $\rho_V >
5\times10^{18}$~$\Omega\cdot$cm. This lower bound is 5 times larger
than the bound given in Ref.~\onlinecite{BLA60}. Note that the current
flowing in LHe due to charge generated by cosmic rays was estimated to
be $I_c \sim 10^{-14}$~A, smaller than the experimentally obtained
limit by two orders of magnitude. The contribution from the natural
radioactivity was also estimated to be $I_r \sim 10^{-14}$~A, based on a measurement of
the radioactivity on the surface of the electrodes.

\section{Summary and outlook}
\label{sec:Summary}
We have designed, constructed, commissioned, and operated an apparatus
for studying electrical breakdown in liquid helium at 0.4 K and
testing electrode materials for the SNS nEDM experiment. It can cool a
6~liter liquid helium volume, which holds a set of electrodes, to
0.4~K. The pressure of the LHe inside this volume can be varied
between the saturated vapor pressure and $\sim$600~torr. High voltage
up to $\pm50$~kV can be applied to each electrode. This apparatus can
be operated to perform two successive cooldowns two weeks apart. Initial
results have demonstrated that it is possible to apply fields exceeding
100~kV/cm in a 1~cm gap between two electropolished stainless steel
electrodes 12~cm in diameter for a wide range of pressures. In
addition, it has been demonstrated that fields exceeding 80~kV/cm can
be applied in a 1~cm gap between two electropolished stainless steel
electrodes 12~cm in diameter with a dielectric insert sandwiched 
between the electrodes with a properly designed groove to accommodate
the insert. In both cases the applicable fields were limited by the
performance of the HV feedlines.

\begin{acknowledgments}
This work was supported by the United States Department of Energy
Office of Nuclear Physics. Development of acrylic-substrate 
electrodes was supported by the Laboratory Directed Research and
Development (LDRD) of Oak Ridge National Laboratory. We gratefully
acknowledge the support of Physics and AOT Divisions as well as the
former LANSCE Division of Los Alamos National Laboratory. We also are
grateful to Brown University Physics Department for making the $^3$He
refrigerator and the cryostat available for this effort. One of the
authors (T.M.I.) expresses his gratitude to Dr. M.~Hardiman of the
University of Sussex for fruitful discussions.
\end{acknowledgments}

\appendix*
\section{Temperature dependence of the pressure of liquid helium in a
  confined container} 

\begin{figure}
\begin{center}
\includegraphics[width=3.5in]{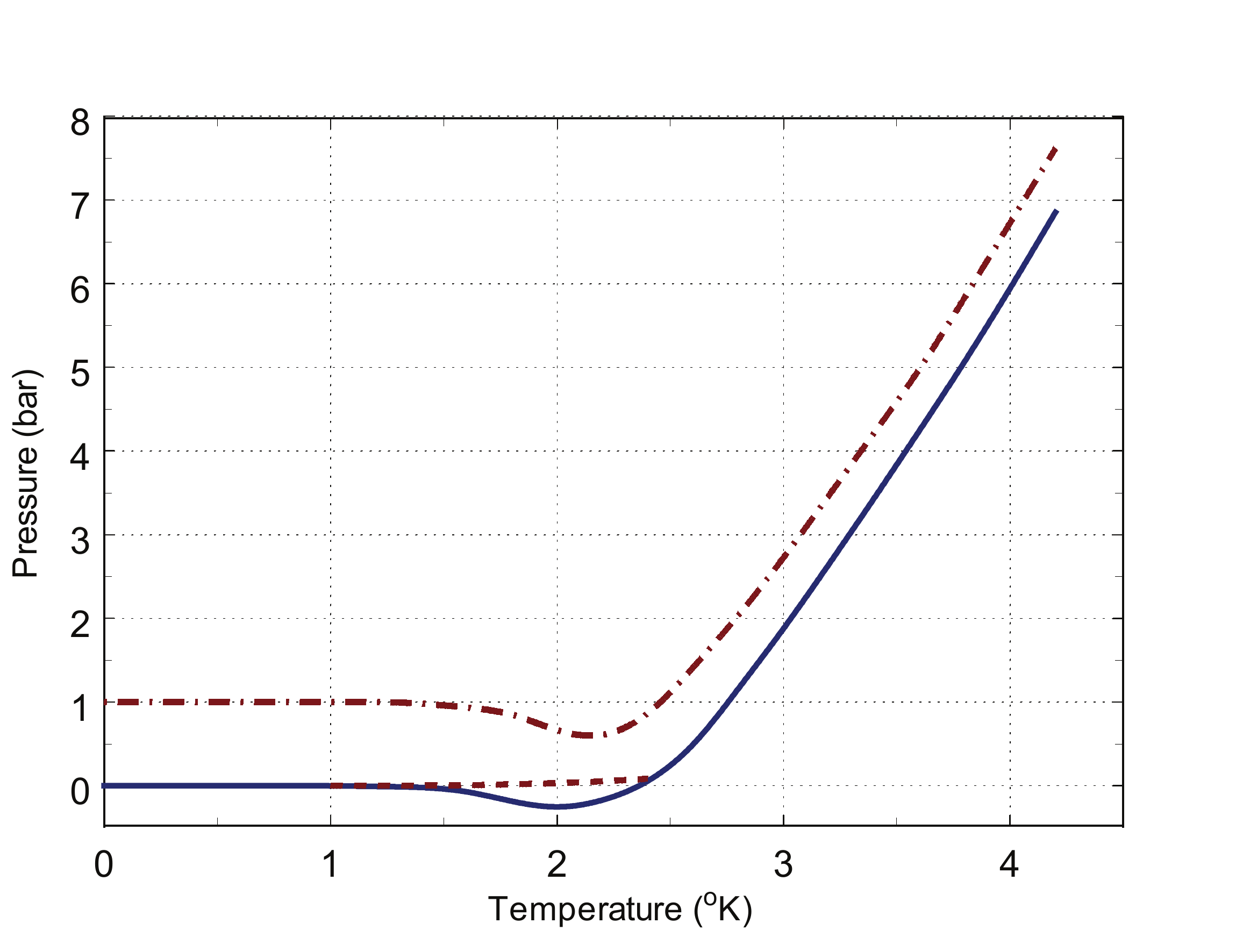}
\end{center}
\caption{Pressure of confined liquid helium plotted as a function of
  the temperature. {\it Solid and dashed lines:} the cell sealed at
  1~K at the saturated vapor pressure (see text for details). {\it
    Dot-dashed line:} the cell sealed at 1~K at a pressure of 1~bar.}
\label{fig:Pconfine}
\end{figure}

Since the thermodynamic properties of liquid helium are well
known\cite{DON98,ELW67,MCC73,MAY76}, it is straightforward to
calculate the temperature dependence of the pressure of a confined
volume of the liquid. The result of such a calculation is shown in
Fig.~\ref{fig:Pconfine} for the liquid sealed at 1 K. The solid curve
is the pressure of the helium sealed at 1~K at the saturated vapor
pressure if it were not to cavitate when the pressure becomes
negative. In the temperature range between 1.15 K and the lambda point
where the thermal expansion of the liquid is negative, the contained
helium develops a vapor phase and the pressure follows the dashed
saturated vapor pressure line in Fig.~\ref{fig:Pconfine}. Above 2.2~K
the pressure within the container rises monotonically with temperature
reaching about 7~bar at 4.2~K. The dot-dashed curve is the pressure of
the helium sealed 1~K at a pressure of 1 bar.

\end{document}